\documentclass[amsmath,amssymb,aps,pre,reprint,showpacs,superscriptaddress]{revtex4-2}
\usepackage{graphicx}
\usepackage{dcolumn}
\usepackage[colorlinks=true,linkcolor=blue,%
pagecolor=blue,citecolor=blue,%
urlcolor=blue,anchorcolor=black,%
bookmarksnumbered=true,%
bookmarksopen=true]{hyperref}
\usepackage[dvipsnames]{xcolor}
\DeclareUnicodeCharacter{2212}{-}
\newcommand{\eqal}[1]{\begin{aligned}#1\end{aligned}}
\newcommand{\autocite}[1]{\cite{#1}}
\newcommand{\mr}[1]{\mathrm{#1}}

\renewcommand{\frac}[2]{\displaystyle{#1 \over #2}}
\newenvironment{myfigure}[3]%
{%
\begin{figure}[htb]
\includegraphics[width=#1\columnwidth]{#2}
\caption{\label{f#2}#3}%
}%
{\end{figure}}%

\begin{document}
\title{Approaching the thermodynamic limit of a bounded one-component plasma}
\author{D.~I.~Zhukhovitskii} \email{dmr@ihed.ras.ru}
\affiliation{Joint Institute of High Temperatures, Russian Academy of Sciences, Izhorskaya 13, Bd.~2, Moscow, 125412 Russia}
\author{E.~E.~Perevoshchikov}
\affiliation{Moscow Institute of Physics and Technology, Institutskiy Pereulok 9, Dolgoprudny, Moscow Region, 141701 Russia}
\affiliation{Joint Institute of High Temperatures, Russian Academy of Sciences, Izhorskaya 13, Bd.~2, Moscow, 125412 Russia}
\date{April 8, 2026}
\begin{abstract}
The classical one-component plasma (OCP) bounded by a spherical surface reflecting ions (BOCP) is studied using molecular dynamics (MD). Simulations performed for a series of sufficiently large BOCPs make it possible to establish the size dependences for the investigated quantities and extrapolate them to the thermodynamic limit. In particular, the total electrostatic energy per ion is estimated in the limit of infinite BOCP in a wide range of the Coulomb coupling parameter $\Gamma$ from 0.03 to 1000 with a relative error of about 0.1\%. The calculated energies are lower by nearly 0.5\% than the modern Monte Carlo (MC) simulation data obtained by different authors at $\Gamma<30$ and almost coincide with the MC results at $\Gamma>175$. We introduce two more converging characteristic energies, the excess interatomic energy and the excess ion–background energy, which enable us to calculate the ionic compressibility factor inaccessible for conventional MC and MD simulations of the OCP with periodic boundary conditions. The derived wide-range ionic equation of state can be recommended for testing OCP simulations with various effective interaction potentials. Based on this equation, we propose an improved cutoff radius for the interionic forces implemented in LAMMPS. We demonstrate that the kinetic and interfacial quantities determined for the classical OCP may be ambiguous. To check this, we perform a simulation of the OCP metastable region and find that its width depends sensitively on the cutoff radius decreasing with the increase of the latter.
\end{abstract}
\pacs{52.27.Gr,52.25.Kn,52.65.−y,36.40.Ei}
\keywords{one-comment plasma, strongly coupled plasma, molecular dynamics, melting}
\maketitle

\section{Introduction}
A system of equal pointlike charges immersed in a uniform neutralizing background of fixed charge density is known as the one-component plasma (OCP) \cite{Baus_1980,Ichimaru_1982,Kholopov_2004,Lieb_2005}. This model is widely explored due to its simplicity that allows one to develop and test models for weakly and, especially, strongly coupled plasmas. Introduction of a rigid background (although some OCP modifications allow for a compressible background) is a convenient way to avoid instability of Coulomb systems. In addition to obvious theoretical importance of OCP models, they have no less important applications in astrophysics because they are appropriate for the interiors of white dwarfs \cite{Ichimaru_1988,Chabrier_1992,Chabrier_1993,Kozhberov_2017,Blouin_2021,Baiko_2022} and neutron stars \cite{Medin_2010}. Such models are also appropriate for liquid alkali metals \cite{Iwamatsu_1986}. Since the confinement field configuration in ion traps is not much different from a parabolic potential, the OCP model is used in the physics of such localized non-neutral plasma objects as ion Coulomb crystals, which are extensively studied both theoretically and experimentally \cite{Wineland_1987,Gilbert_1988,Waki_1992,Itano_1998,Dubin_1999,Bollinger_2000,Huang_2002,Mortensen_2006,Mavadia_2013,Yan_2016}. In these studies, a crystallized core was observed and simulated, e.g., in spherical plasmas with more than $2\times10^5$ ions, the formation of bcc crystals occupying the inner quarter of the plasma diameter was detected \cite{Bollinger_2000}.

Another application of the OCP model is the Coulomb cluster (Coulomb ball) in dusty plasmas \cite{Arp_2004,Arp_2005,K_ding_2006,Totsuji_2006,Apolinario_2014} as well as the quasi-homogeneous dusty plasma \cite{Zhukhovitskii_2017,Huang_2023}. Besides the classical OCP, quantum OCP is widely studied \cite{Pokrant_1977,Ceperley_1978}, especially for astrophysics \cite{Baiko_2000,Chugunov_2003,Chugunov_2005}. The screened Coulomb (or Yukawa) systems are being studied no less intensively \cite{Hamaguchi_1994,Farouki_1994,Hamaguchi_1996,Caillol_2000_2,Caillol_2000_3,Bonitz_2006,Khrapak_2014}. In some studies, properties of the classical OCP are deduced in the limiting case of the Yukawa system with infinite Debye length. In the theoretical studies of spherical Coulomb crystals \cite{Totsuji_2002,Hasse_2003} performed for extremely low temperatures and the system sizes up to $1.2\times10^5$, the cluster interior with crystal ordering surrounded by a few shells on the outside was found.

For the classical OCP, the most studied quantities are the internal energy, the Helmholtz free energy, and the equation of state (EOS). Investigations are performed using analytical methods \cite{Young_1991,Brilliantov_1998,Chabrier_1998,Caillol_1999,Caillol_1999_2,Brilliantov_2002,Khrapak_2014,Khrapak_2016}, Monte Carlo (MC) \cite{Brush_1966,Hansen_1973,Slattery_1980,Slattery_1982,Ogata_1987,Caillol_1999_3,Caillol_2010,Demyanov_2022}, and molecular dynamics (MD) simulation \cite{Farouki_1993}. Extrapolation of the MC data to infinite number of ions allows one to obtain the most accurate estimate of the system electrostatic energy (with the relative error about 0.1\% and better). Extensively studied is melting of the OCP \cite{Pollock_1973,Weeks_1981,Nordholm_1984,Lee_1988,Likos_1992}, the liquidlike to gaslike dynamical crossover \cite{Huang_2023,Yu_2024,Khrapak_2025}; the OCP transport properties are also of interest \cite{Khrapak_2025}.

Conventional conditions used in OCP simulations are periodic boundary conditions (PBC) \cite{Allen_2017}, which are inevitable in the case of long-range Coulomb forces because they substantially minimize a strong effect of the system boundaries. PBC imply that the simulated system consists of a cubic supercell of the edge length $L$  containing the ions along with an infinite lattice of its periodic images. Each image cell is obtained by translating the original supercell by $sL$ along one or more coordinate axes, where $s=\pm1,\pm2,\pm3,\ldots$ Thus, an ion in the supercell with the radius-vector ${\mathbf{r}}$ has image ions at positions ${\mathbf{r}} + {\mathbf{s}}L$, where $\mathbf{s}$ is a vector of integers, and $\mathbf{s}\ne0$. For such a system, the electrostatic energy is represented as a conditionally converging series. Since this series converges very slowly, a method based on summation in reciprocal space developed by Ewald \cite{Ewald_1921} (see also \cite{Nijboer_1957}) is used along with the modifications increasing its efficiency (e.g., \cite{Karasawa_1989,Yakub_2003,Demyanov_2024}). For the PBC (or bulk) OCP, the only convergent energy characteristic is the total electrostatic energy per ion; at least, no other converging quantity can be obtained directly from the Ewald summation. In contrast, finite-size systems with the finite number of ions $N$ and an explicit boundary offer a possibility to search for such characteristics because each of them becomes formally finite and can be calculated directly by summation of individual Coulomb interactions, which does not require the Ewald summation. As is known, under certain conditions, the Coulomb system reaches a thermodynamic limit at infinite size. Therefore, if it is possible to establish a size dependence for the quantity of interest at sufficiently large $N$, then a reliable extrapolation to the thermodynamic limit can be derived.

In our recent study \cite{Zhukhovitskii_2024}, we investigated large Coulomb clusters with a finite spherical neutralizing background and demonstrated that for $N>2500$, a quasi-homogeneous cluster core can be found that can be treated as a satisfactory approach to the bulk system with compatible characteristics. Here, our results are very close to those obtained in \cite{Hasse_2003}. In particular, we have also found that the core has a polycrystalline structure represented by bcc, fcc, and hcp crystallites. In contrast to \cite{Totsuji_2002,Hasse_2003}, where the Coulomb coupling parameter $\Gamma$ was set to $10^5\mbox{--}10^6$, we considered much lower $\Gamma \gtrsim 200$, which corresponded to the region of cluster melting crossover, and investigated the dependence of the spherical shell structure and the core size on $\Gamma$. In addition, we showed that the ionic compressibility factor could be expressed in terms of the energies of interionic interaction and their interaction with the cluster background. However, we only obtained a crude estimate of $Z_i$ at $\Gamma=500$ and noticed that the ion evaporation breaks noticeably both the system stability and its local quasineutrality at $\Gamma<200$.

In this work, we consider a bounded OCP (BOCP), which is free from these shortcomings, with a rigid boundary that reflects the ions specularly and is situated at the surface of the background sphere. In this model, it is possible to investigate the entire range of $\Gamma$ from weakly to strongly coupled systems and to determine the size dependence for the quantities of interest. In addition to the conventional total electrostatic energy, we introduce two more energy quantities, the excess interatomic and ion--background electrostatic energy, and show them to be convergent in the thermodynamic limit. We also deduce the ionic pressure, i.e., the pressure of the ionic subsystem, as a linear combination of these energies. This enables us to construct the ionic EOS verified by determination of the ionic virial, which must hold for ordinary OCP.

Yet another problem associated with the reciprocal space summation is an explicit dependence of the resulting interionic effective potential on the simulation cell length, which proves to be the potential decay length. This means that the Ewald potential, although decaying much more rapidly than the Coulomb one, is still a slowly decaying one. This is a source of ambiguity in taking into account the interaction of ions with their periodic images, which was noted already in the early study \cite{Brush_1966}. The most sensitive to this ambiguity is MD simulation, so that alternative methods for summation in $k${}-space were proposed such as that in \cite{Karasawa_1989}, in which the contribution from neighboring ions at distances less than a properly chosen Coulomb cutoff radius $r_c$ is included directly while that of distant ions is calculated by the $k${}-space summation. Onegin \textit{et al.} \cite{Onegin_2024} were the first who discovered that the virial determined in the MD simulation of OCP depends on $r_c$, and moreover, this virial disagrees strongly with the pressure that is derived from the system potential energy. Again, the \textit{ab initio} BOCP simulation does not encounter such problem, and we can compare the ionic pressure from our simulation with that for OCP. Thus, we propose to “fix” the MD simulations for OCP by taking $r_c$ as a function of $\Gamma$ when our $Z_i$ is used as reference data. As an illustration, we demonstrate that a different choice of $r_c$ changes significantly the width of metastable region of fluid--solid phase transition thus indicating the change in interfacial surface tension.

In Sec.~\ref{bkm:RefHeadingToc71623386547914}, we introduce the main BOCP energy characteristics and explore them in limiting cases; in Sec.~\ref{bkm:RefNumPara97513304080641}, we derive the ionic EOS as well as its excess ionic Helmholtz free energy. The method of MD simulation along with its justification and typical parameters is presented in Sec.~\ref{bkm:RefNumPara11801874321070}. In Sec.~\ref{bkm:RefNumPara9096183868828}, the MD results for BOCP are analyzed and discussed. Secs.~\ref{bkm:RefNumPara55813304080641} and \ref{bkm:RefHeadingToc409121406508568} are devoted to the MD simulation of the OCP with PBC: the details of simulation method are discussed in Sec.~\ref{bkm:RefNumPara55813304080641}; in Sec.~\ref{bkm:RefHeadingToc409121406508568}, the recommended dependence $r_c(\Gamma)$ is presented. The effect of this dependence on the kinetics of fluid--solid phase transition is examined in Sec.~\ref{bkm:RefHeadingToc183112903921931}.

\section{Theoretical background}
\subsection[Energy characteristics of the bounded one{}-component plasma]{Energy characteristics of the bounded one-component plasma}
\label{bkm:RefHeadingToc71623386547914}We treat a system of $N$ particles (ions) with the mass $m$ and charge $ze$, where $e$ is the elementary charge, on a uniform background of the total charge $- zeN$ bounded by a sphere of the radius $R$. To prevent ion evaporation at high temperature, i.e., departure of the faster ions to infinity, the system is bounded by a rigid spherical boundary, from which the ions are reflected specularly. Since we develop an asymptotic model of the OCP, its radius $R_b$ must coincide with that of the background; otherwise, the system cannot be homogeneous even at low $\Gamma$. However, finiteness of the MD integration time step requires that $R_b$ is somewhat less than $R$ (Sec.~\ref{bkm:RefNumPara11801874321070}).

In what follows, we will term such system the BOCP. In contrast to OCP, here, another key parameter $N$ is added to the Coulomb coupling parameter $\Gamma = \epsilon /T$, where $\epsilon = {z^2}{e^2}/{r_s}$ is the characteristic electrostatic energy per ion, ${r_s} = {(3/4\pi {\bar n})^{1/3}}= R{N^{ - 1/3}}$ is the ion sphere radius, $\bar n = 3N/4\pi {R^3}$ is the average ion number density, $T$ is the temperature in energy units; BOCP is assumed to be in the equilibrium state, and $N$ is sufficiently large. Hereafter, we use the Coulomb units by setting $r_s=m=\epsilon=ze=1$. Then the unit time is ${\tau _0} = {r_s}{(m/\epsilon )^{1/2}} = {(mr_s^2/\Gamma T)^{1/2}}$, which coincides with the inverse oscillation frequency of an ion in the ion sphere. In the Coulomb units, $T=1/\Gamma$, $R=N^{1/3}$, and $\bar{n}=3/4\pi$.

The total electrostatic energy of the BOCP per ion is
\begin{equation}\label{seq:refEquation0}
u(\Gamma,N) = {u_p} + {u_b} + {u_c},
\end{equation}
where
\begin{equation}\label{seq:refEquation1}
{u_p} = \frac{1}{N}\sum\limits_{k < l}\frac{1}{r_{kl}} = \frac{1}{2N}\iint\limits_{V_{R}}{\frac{{n({\mathbf{r}}_{1}')n({\mathbf{r}}_{2}')}}{{{r'_{12}}}}\,d{\mathbf{r}}_{1}'d{\mathbf{r}}_{2}' }
\end{equation}
is the interionic interaction energy per ion, $r_{12} = |{\mathbf{r}}_{1} - {\mathbf{r}}_{2}|$, $n({\mathbf{r}}') = \sum\nolimits_{k = 1}^N {\delta ({\mathbf{r}}' - {{\mathbf{r}}_k})} $ is the ion number density, ${\mathbf{r}}_k$ is the radius-vector of the $k$th ion, and the integration is performed in the background volume $V_R$;
\begin{equation}\label{seq:refEquation2}
\eqal{
{u_b} & = \frac{1}{{2N}}\sum\limits_{k = 1}^N {r_k^2}  - \frac{5}{2}{u_{c}} \\ 
& = \frac{1}{{2N}}\int\limits_{{V_R}} {{{r'}^2}n({\mathbf{r}'})} \,d{\mathbf{r}'} - \frac{5}{2}{u_{c}}
}
\end{equation}
is the energy of interaction between the ions and background per ion, and
\begin{equation}\label{seq:refEquation3}
{u_c} = \frac{3}{5}{N^{2/3}}
\end{equation}
is the background energy per ion. Equation (\ref{seq:refEquation2}) implies that the origin of the coordinate system coincides with the background center. Although $u$ converges at $N\to\infty$, $u_p$ and $u_b$ diverge as $N^{2/3}$; however, one can define renormalized quantities that will be termed the excess interatomic and ion--background electrostatic energy
\begin{equation}\label{seq:refEquation4}
{u_{p\mr{ex}}} = {u_p} - {u_c},\quad {u_{b\mr{ex}}} = {u_b} + 2{u_c},
\end{equation}
respectively, such that the total energy $u={u_{p\mr{ex}}}+{u_{b\mr{ex}}}$ coincides with the excess total electrostatic energy.

We will demonstrate convergence of $u_{p\mr{ex}}$ and $u_{b\mr{ex}}$ in the cases of weakly and strongly coupled systems. If the system is gaslike, and the ternary and higher-order ion correlations can be neglected, then
\begin{equation}
\left\langle {n({{\mathbf{r}}_1})n({{\mathbf{r}}_2})} \right\rangle \simeq \left\langle  {n(0)n(r_{12})} \right\rangle  = {\left( {\frac{3}{{4\pi }}} \right)^2}{f_p}(r_{12}),
\end{equation}
where brackets denote time averaging,
\begin{equation}\label{seq:refEquation6}
{f_p}(r) = \frac{1}{{3{r^2}}}\frac{{d{N_r}}}{{dr}}
\end{equation}
is the radial distribution function, and $N_r$ is the average number of ions within a sphere of the radius $r$ around an ion. Similarly, $\left\langle {n({\mathbf{r}})} \right\rangle  = \left\langle {n(r)} \right\rangle  = (3/4\pi ){f_c}(r)$, where
\begin{equation}\label{seq:refEquation7}
{f_c}(r) = \frac{1}{{3{r^2}}}\frac{{d{N_c}}}{{dr}}
\end{equation}
is the ion density distribution and $N_c$ is the average number of ions at the distance not larger than $r$ from the background center. Then from Eqs.\ (\ref{seq:refEquation1}) and (\ref{seq:refEquation2}) we obtain
\begin{equation}\label{seq:refEquation8}
\eqal{
\left\langle{{u_{p\mr{ex}}}}\right\rangle  & \simeq \frac{1}{2N}{\left( {\frac{3}{{4\pi }}} \right)^2}\iint\limits_{V_{R}}{\frac{{f_p}(r_{1}){f_p}(r_{2})-1}{{r_{12}}}\,d{\mathbf{r}}_{1}d{\mathbf{r}}_{2} }\\ 
& = \frac{9}{{4N}}\int\limits_0^R {\int\limits_0^R {\left[ {{f_p}({r_1}){f_p}({r_2}) - 1} \right] }}\\
& \times \left( {{r_1} + {r_2} - \left| {{r_1} - {r_2}} \right|} \right){r_1}{r_2}\,d{r_1}d{r_2}
}
\end{equation}
and
\begin{equation}\label{seq:refEquation9}
\left\langle{{u_{b\mr{ex}}}}\right\rangle = \frac{3}{{2N}}\int\limits_0^R {{r^4}\left[ {{f_c}(r) - 1} \right]\,dr} .
\end{equation}
Since the ion correlations decay at $\Gamma \to 0$, then ${f_p},{f_c} \to 1$, and from Eqs.\ (\ref{seq:refEquation8}) and (\ref{seq:refEquation9}), it follows that
\begin{equation}
\mathop {\lim }\limits_{\Gamma  \to 0 } \left\langle{{u_{p\mr{ex}}}}\right\rangle  = \mathop {\lim }\limits_{\Gamma  \to 0 } \left\langle{{u_{b\mr{ex}}}}\right\rangle  = 0.
\end{equation}
Thus, in the thermodynamic limit, we obtain an obvious result: all excess electrostatic energies of an ideal gas vanish:
\begin{equation}\label{seq:refEquation11}
\mathop {\lim }\limits_{N \to \infty } \mathop {\lim }\limits_{\Gamma  \to 0 }
\left\langle{{u_{p\mr{ex}}}}\right\rangle 
 = \mathop {\lim }\limits_{N \to \infty } \mathop {\lim }\limits_{\Gamma  \to 0 } \left\langle{{u_{b\mr{ex}}}}\right\rangle
=u_{\infty} = 0,
\end{equation}
where $u_{\infty} = \mathop {\lim }\limits_{N \to \infty } \left\langle{u}\right\rangle $. Note that if we treat separately the system of ions in the field of background then the ionic potential energy $u_i = {u_p} + {u_b} $ diverges as $N\to\infty$. However, the excess ionic potential energy $u_i + u_c={u_{p\mr{ex}}}+{u_{b\mr{ex}}}$, which coincides with the excess total electrostatic energy, does converge.

Consider the case $\Gamma \to \infty$, in which the ion sphere model (see, e.g., \cite{Ichimaru_1982}) proved to be most effective for the Coulomb clusters \cite{Shpil_ko_2023,Zhukhovitskii_2024}. This model implies that the entire system is divided into spherical overlapping spheres of the radius $r_s$, in which single ions reside, and all interparticle and particle--background interaction can be reduced to the interaction of an ion with the compensating background of its sphere. Thus, the net force acting on an ion is simply ${\mathbf{f}} = - {\mathbf{r}}$, where the origin of the coordinate system is shifted to the center of an ion sphere and we omit subscripts due to identity of the ions. Based on the virial theorem \cite{Landau_Statphys} and an obvious equality for a harmonic oscillator $\left\langle {{r^2}} \right\rangle  = \left\langle {{v^2}} \right\rangle  = 3/\Gamma $, we can write the ionic compressibility factor as \cite{Zhukhovitskii_2024}
\begin{equation}\label{seq:refEquation12}
{Z_i} = 1 + \frac{\Gamma }{{3N}}\sum\limits_{i = 1}^N {\left\langle {{\mathbf{rf}}} \right\rangle }  = 1 - \frac{\Gamma }{3}\left\langle {{r^2}} \right\rangle = 0,
\end{equation}
i.e., the ionic compressibility factor along with the ionic pressure $p_i$ vanish at $\Gamma \to \infty$.

We will use the virial theorem one more time to express the ion pressure and corresponding compressibility factor in terms of the ion subsystem potential and kinetic energy. However, if we assume that the charge of neutralizing background of BOCP is fixed, $r_s =\mr{const}$, then $u_{p}$ (Eq.\ref{seq:refEquation1}) is a homogeneous function of order $-1$; $u_{b}+5u_c/2$, of order 2 (Eq.\ (\ref{seq:refEquation2})); and $u_c$, of order 0. Bearing in mind that if the function $\varphi$ is a sum of homogeneous functions of order $k$, $\varphi (x) = \sum\nolimits_k {{\psi _k}(x)}$, then
\begin{equation}
\frac{{d\varphi }}{{dx}} = \sum\limits_k {\frac{{k{\psi _k}(x)}}{x}} .
\end{equation}
Thus, one can modify the virial theorem for the ions in the field of background as
\begin{equation}\label{seq:refEquation14}
3{p_i}{V_R} = \frac{3}{\Gamma } +  {u_p} - 2{u_b} - 5{u_c}  ,
\end{equation}
where $3/\Gamma$ is the kinetic energy of an ion, whence it follows that \cite{Zhukhovitskii_2024}
\begin{equation}\label{seq:refEquation15}
{Z_i} = \frac{{{p_i}{V_R}}}{{NT}} = 1 + \frac{\Gamma }{3}({u_{p\mr{ex}}} - 2{u_{b\mr{ex}}}).
\end{equation}
Here and in what follows, the energy quantities will be implied time averaged, and corresponding notation will be omitted.

From Eqs.\ (\ref{seq:refEquation12}) and (\ref{seq:refEquation15}),
\begin{equation}
\mathop {\lim }\limits_{\Gamma  \to \infty }  {\frac{{{Z_i} - 1}}{\Gamma }}  = \mathop {\lim }\limits_{\Gamma  \to \infty } ({u_{p\mr{ex}}} - 2{u_{b\mr{ex}}}) = 0.
\end{equation}
At the same time, $\mathop {\lim }\limits_{N  \to \infty }\mathop {\lim }\limits_{\Gamma  \to \infty } u = \mathop {\lim }\limits_{N  \to \infty }\mathop {\lim }\limits_{\Gamma  \to \infty } ({u_{p\mr{ex}}} + {u_{b\mr{ex}}}) = {u_0}$, where $u_0=-0.8959293$ is the Madelung constant for the bcc lattice. Hence,
\begin{equation}\label{seq:refEquation17}
\mathop {\lim }\limits_{N  \to \infty }\mathop {\lim }\limits_{\Gamma  \to \infty }{u_{p\mr{ex}}} =  \frac{2}{3}{u_0},\;\; \mathop {\lim }\limits_{N  \to \infty }\mathop {\lim }\limits_{\Gamma  \to \infty }{u_{b\mr{ex}}} =  \frac{1}{3}{u_0}.
\end{equation}
Thus, we have demonstrated that the quantities (\ref{seq:refEquation4}) have thermodynamic limits at low and high $\Gamma$, which is indicative of the fact that these limits exist at arbitrary $\Gamma$ defining the ionic pressure for the bulk OCP.

Since an objective of this paper is investigation of the OCP properties, the size dependence $u(\Gamma,N)$ (\ref{seq:refEquation0}) is of interest. As it was demonstrated in \cite{Zhukhovitskii_2024}, all quantities derived from the BOCP energy characteristics depend on $N$ very weakly, therefore, we adopt $\ln N$ as an argument. The simplest and most general form of the size dependence is then the expansion in powers of shifted $\ln N$, where we retain only two terms:
\begin{equation}\label{seq:refEquation18}
u(\Gamma,N) = {u_\infty }(\Gamma) + \frac{{c_1}(\Gamma)}{{\ln N + {c_2}(\Gamma)}}.
\end{equation}
Here, $u_{\infty}(\Gamma)$, $c_1$, and $c_2$ are the parameters to be defined from fitting the MD simulation results (Sec.~\ref{bkm:RefNumPara9096183868828}).

\subsection{Equations of state and thermodynamic functions}
\label{bkm:RefNumPara97513304080641}In Sec.~\ref{bkm:RefHeadingToc71623386547914}, we have explored the ionic subsystem and derived EOS (\ref{seq:refEquation15}) for it. Below, we will obtain the EOS for the entire BOCP, which is supposed to include a contribution from the background, and demonstrate that the pressure for the entire BOCP is principally different from the ionic pressure. This requires some physical model of the background. We consider the simplest model of the background as a system of a large number $\bar N \to \infty$ of small charged pseudo-particles of the charge ${\bar z}e$ uniformly distributed in the volume $V_R$ such that $\bar N \bar z + Nz =0$ \cite{Weeks_1981}. Here, the quantities related to the pseudo-particles are marked with bars. In this limit, it is obvious that both the pseudo-particle--pseudo-particle and pseudo-particle--ion coupling parameters, $\Gamma {(N/\bar N)^{5/3}}$ and $\Gamma {(N/\bar N)^{2/3}}$, respectively, vanish, and the background electrostatic energy is defined by Eq.\ (\ref{seq:refEquation3}). Now, we have to treat the system potential energy as a homogeneous function of the order $-1$ in the coordinates and write the EOS for entire system based on the virial theorem as
\begin{equation}\label{seq:refEquation19}
p = \frac{N}{{{V_R}}}\left( {T + \frac{{2\bar K}}{{3N}} - \frac{u}{3}} \right),
\end{equation}
where $p$ is the system pressure and $\bar K = (3/2)\bar NT$ is the pseudo-particle kinetic energy. As is well known, the totally neutral system of classical Coulomb charges tends to collapse, and the main background function is to stabilize the entire system. This is reached by the ideal gas pressure of the pseudo-particles, which tends to infinity as $\bar N \to \infty$. However, one can renormalize this pressure by introducing the excess pressure $p \to p - 2\bar K/3{V_R}$ \cite{Weeks_1981}, so that the compressibility factor is reduced to
\begin{equation}\label{seq:refEquation20}
Z = 1 + \frac{{\Gamma u}}{3}
\end{equation}
(cf.\ \cite{Brush_1966}). In terms of virials, Eq.\ (\ref{seq:refEquation19}) with $\bar K=0$ can be written as
\begin{equation}\label{seq:refEquation21}
p = \frac{N}{{{V_R}}}\left( {T + \frac{1}{{3N}}\sum\limits_{i = 1}^N {\left\langle {{{\mathbf{r}}_i}{{\mathbf{f}}_i}} \right\rangle }  + \frac{1}{{3N}}\sum\limits_{j = 1}^{\bar N} {\left\langle {{{{\mathbf{\bar r}}}_j}{{{\mathbf{\bar f}}}_j}} \right\rangle } } \right),
\end{equation}
whereas the expression for $p_i$ (\ref{seq:refEquation14}) corresponds to
\begin{equation}
p_i = \frac{N}{{{V_R}}}\left( {T + \frac{1}{{3N}}\sum\limits_{i = 1}^N {\left\langle {{{\mathbf{r}}_i}{{\mathbf{f}}_i}} \right\rangle } } \right)
\end{equation}
and (\ref{seq:refEquation15}), to
\begin{equation}\label{seq:refEquation23}
{Z_i} = 1 + \frac{\Gamma }{{3N}}\sum\limits_{i = 1}^N {\left\langle {{{\mathbf{r}}_i}{{\mathbf{f}}_i}} \right\rangle } .
\end{equation}
The pressures $p$ and $p_i$ and corresponding compressibility factors $Z$ and $Z_i$ differ by the third term in parentheses on the right-hand side of (\ref{seq:refEquation21}), which cannot be determined in any simulation. Therefore, although the pressure (\ref{seq:refEquation21}) must be fully compatible with its thermodynamic definition because it leads to (\ref{seq:refEquation20}), it cannot be obtained from simulations. The resulting incompatibility of entirely different quantities, $p$ and $p_i$ is worth mentioning. It is $p_i$ that is actually determined in both MD and MC simulations, therefore, $p_i$ cannot be compared with the thermodynamically defined system pressure $p$ regardless of whether it is renormalized, despite the fact that both the thermodynamic approach and virial theorem are suitable for the treated system.

If we treat the ions in the field of background as a separate system, we can define the excess internal energy of ions $\varepsilon_{\mr{ex}}(\Gamma,N)$ per ion, which coincides with the excess internal energy of the entire system (Sec.~\ref{bkm:RefHeadingToc71623386547914}). If we choose $\Gamma=\infty$ as a reference state, then
\begin{equation}\label{seq:refEquation24}
\varepsilon_{\mr{ex}}(\Gamma,N)=u(\Gamma,N)-u(\infty,N).
\end{equation}
In the thermodynamic limit, $\varepsilon_{\mr{ex}}(\Gamma)=u_{\infty}(\Gamma)-u_0$. At $\Gamma  > {\Gamma _\mr{m}}$, where $\Gamma={\Gamma _\mr{m}}$ is the melting point of OCP, the harmonic lattice model supplemented by anharmonicity correction is applicable \cite{Itoh_1980,Chugunov_2005}, which yields
\begin{equation}\label{seq:refEquation25}
\varepsilon_{\rm{ex}}(\Gamma) = \frac{3}{2\Gamma} + \frac{\beta_2}{\Gamma^2}
\end{equation}
with the coefficient $\beta_2$ defining anharmonicity. The thermodynamic limit of the excess Helmholtz free energy per ion of the entire system, which is equal to that of the ionic subsystem, can be derived from $\varepsilon_{\rm{ex}}(\Gamma)$ (i.e., from $u_\infty(\Gamma)$, see, e.g., \cite{Chabrier_1998}):
\begin{equation}\label{seq:refEquation26}
{f_{\mr{ex}}}(\Gamma) =  - \int\limits_\Gamma ^\infty  {\frac{{{\varepsilon _{\mr{ex}}}(\Gamma ')}}{{\Gamma '}}} \,d\Gamma ' .
\end{equation}
Then at the melting point, we obtain from (\ref{seq:refEquation25}) and (\ref{seq:refEquation26})
\begin{equation}\label{seq:refEquation27}
{f_{\mr{ex}}}({\Gamma _\mr{m}}) =  - \frac{3}{{2{\Gamma _\mr{m}}}} - \frac{\beta _{2} }{{2\Gamma _\mr{m}^2}}.
\end{equation}
\section{MD simulation for BOCP}
\subsection{Simulation method}
\label{bkm:RefNumPara11801874321070}An advantage of the BOCP simulation is the absence of diverging quantities due to finiteness of $N$, which makes it possible to determine all quantities of interest including $u$, $u_p$, and $u_b$. In contrast, calculation of the total potential energy $u$ for the OCP with PBC requires application of special summation methods such as the Ewald potential \cite{Ewald_1921,Brush_1966}, angular-averaged Ewald potential \cite{Demyanov_2022}, particle-particle particle-mesh method (PPPM) \cite{Hockney_2021}, etc., while for BOCP, this quantity can be determined \textit{ab initio} from direct summation of the interparticle Coulomb energies and the particle--background energy. Moreover, since no summation method for $u_{p\mr{ex}}$ and $u_{b\mr{ex}}$ for the OCP has been developed, BOCP simulation remains a unique method for the estimation of these quantities. Obvious shortcomings of the BOCP simulation are enormous computational efforts when no acceleration procedure is possible and the necessity to simulate a whole spectrum of the BOCP sizes to enable the extrapolation to the thermodynamic limit.

MD simulation of ions motion within the background sphere is performed. Introduction of the dimensionless dynamic quantities by replacing the time $t$ $\to t/{\tau _0}$, the radius-vector of the $i$th ion ${{\mathbf{r}}_i} \to {{\mathbf{r}}_i}/{r_s}$, and the force acting on this ion ${{\mathbf{f}}_i} \to ({r_s}/\varepsilon ){{\mathbf{f}}_i} = \left( {\tau _0^2/m{r_s}} \right){{\mathbf{f}}_i}$ makes it possible to write the equation of the ion motion in the form \cite{Zhukhovitskii_2024}
\begin{equation}\label{seq:refEquation28}
{{\mathbf{\ddot r}}_i} = {{\mathbf{f}}_{zi}} + {{\mathbf{f}}_{bi}} + {{\mathbf{f}}_{Li}},
\end{equation}
where
\begin{equation}
{{\mathbf{f}}_{zi}} = \sum\limits_{j \ne i} {\frac{{{\mathbf{r}}_i} - {{\mathbf{r}}_j}}{r _{ij}^3}} 
\end{equation}
is the force of interparticle interaction,
\begin{equation}
{{\mathbf{f}}_{bi}} = \left\{ {\begin{array}{cc}
{ - {{\mathbf{r}}_i},}&{\mbox{ }{r_i} \le R,}\\
{ - N\frac{{{{\mathbf{r}}_i}}}{{r_i^3}},}&{\mbox{ }{r_i} > R}
\end{array}} \right.
\end{equation}
is the force of attraction to the background, and
\begin{equation}
{{\mathbf{f}}_{Li}} = - \gamma {{\mathbf{\dot r}}_i} + {{\mathbf{f}}_{\mathrm{st}i}}
\end{equation}
is the force acting from the Langevin thermostat. Here, $\gamma$ is the particle friction coefficient, ${{\mathbf{f}}_{\mathrm{st}i}}$ is a random force with the Gaussian distribution such that
\begin{equation}\label{seq:refEquation32}
\left\langle {f_{\mr{st}i}^2} \right\rangle  = \frac{{2\gamma \left\langle {{v^2}} \right\rangle }}{{{\tau _{\mr{st}}}}} = \frac{{6\gamma }}{{\Gamma {\tau _{\mr{st}}}}},
\end{equation}
where ${\tau _{\mathrm{st}}}$ is the autocorrelation decay time, which was set equal to the time integration step $\tau$, $\left\langle {{v^2}} \right\rangle  \equiv \left\langle {v_i^2} \right\rangle $ is the mean squared ion velocity, which is independent of $i$ at equilibrium, and we take into account that in equilibrium, $\left\langle {{v^2}} \right\rangle = 3T=3/\Gamma$.

Contact with the thermostat in the course of simulation reduces sharply the system equilibration time and ensures a very small deviation of $\left\langle {{v^2}} \right\rangle /3$ from the desired $T$ thus decreasing substantially the computation time. However, the question arises whether the inclusion of ${\mathbf{f}}_{Li}$ in the total force can distort the virial. Below, we demonstrate that no distortion can be observed because $\sum\nolimits_{i = 1}^N {{{\mathbf{r}}_i} {{\mathbf{f}}_{Li}}} = 0$. First, consider the case of ideal gas ($\Gamma \to 0$), \textrm{in which case }Eq.\ (\ref{seq:refEquation28}) can be rewritten as
\begin{equation}\label{seq:refEquation33}
{{\mathbf{\ddot r}}_i} =  - \gamma {{\mathbf{\dot r}}_i} + {{\mathbf{f}}_{\mr{st}i}}.
\end{equation}
We multiply both sides of (\ref{seq:refEquation33}) by ${\mathbf{r}}_i$, average this equation, and take into account that $\left\langle {{{\mathbf{r}}_i} {{{\mathbf{\ddot r}}}_i}} \right\rangle  = d\left\langle {{{\mathbf{r}}_i} {{{\mathbf{\dot r}}}_i}} \right\rangle /dt - \left\langle {{v^2}} \right\rangle $ and
\begin{equation}
\frac{d}{{dt}}\left\langle {{{\mathbf{r}}_i} {{{\mathbf{\dot r}}}_i}} \right\rangle  = \frac{1}{2}\frac{{{d^2}}}{{d{t^2}}}\left\langle {r_i^2} \right\rangle  = \frac{D}{2}\frac{{{d^2}t}}{{d{t^2}}} = 0,
\end{equation}
where we imply that an ion is involved in the Brownian motion when $\left\langle {r_i^2} \right\rangle = {\mathcal D}t$, where, as it follows from (\ref{seq:refEquation32}), ${\mathcal D} = {\tau _{\mr{st}}}\left\langle {{v^2}} \right\rangle $. Bearing this in mind we infer from Eqs.\ (\ref{seq:refEquation32}) and (\ref{seq:refEquation33}) $\left\langle {{{\mathbf{r}}_i}{{\mathbf{f}}_{\mr{st}i}}} \right\rangle  = 0$, so that the virial vanishes:
\begin{equation}
\sum\limits_{i = 1}^\infty  {\left\langle {{{\mathbf{r}}_i}{{\mathbf{f}}_{Li}}} \right\rangle }  =  - \gamma \sum\limits_{i = 1}^\infty  {\left\langle {{{\mathbf{r}}_i}{{{\mathbf{\dot r}}}_i}} \right\rangle }  + \sum\limits_{i = 1}^\infty  {\left\langle {{{\mathbf{r}}_i}{{\mathbf{f}}_{\mr{st}}}} \right\rangle }  = 0.
\end{equation}

Next, consider the case of a strongly coupled system, $\Gamma \to \infty$, for which we use again the ion sphere model (Sec.~\ref{bkm:RefHeadingToc71623386547914}). Now the ion equation of motion includes the electrostatic force ${\mathbf{f}} =  - {\mathbf{r}}$: ${\mathbf{\ddot r}} =  - {\mathbf{r}} - \gamma {\mathbf{\dot r}} + {{\mathbf{f}}_{\mr{st}}}$, where we omit subscripts. In contrast to the case of ideal gas treated above, an ion in a strongly coupled system is involved in the Ornstein--Uhlenbeck process in its microscopic sphere, so that $\left\langle {{\mathbf{r}}{\mathbf{\dot r}}} \right\rangle  = (1/2) d\left\langle {{r^2}} \right\rangle /dt = 0$, and therefore, $\left\langle {{\mathbf{r}}{\mathbf{\ddot r}}} \right\rangle  =  - \left\langle {{v^2}} \right\rangle $. We multiply both sides of the ion equation of motion by ${\mathbf{r}}$ with due regard for the equality $\left\langle {{v^2}} \right\rangle  = \left\langle {{r^2}} \right\rangle $ valid at equilibrium to derive the same result as above: ${\mathbf{r}}{{\mathbf{f}}_L} = 0$ \cite{Zhukhovitskii_2024}.

In the intermediate case of moderate $\Gamma$, the ion is involved in both the Brownian and Ornstein--Uhlenbeck processes, i.e., its motion is characterized by two time scales. Such process is quite similar to that observed in study \cite{Zhukhovitskii_2024_2}. Test runs showed that irrespective of $\Gamma$, all results were absolutely insensitive to the choice of $\gamma$, which confirms the conclusions made above (note that this must be true solely at equilibrium). For all BOCP simulations, we take $\gamma \simeq 0.09$ from earlier studies \cite{Zhukhovitskii_2017,Shpil_ko_2023,Zhukhovitskii_2024}.

We apply the specular reflection boundary conditions for the equation of motion (\ref{seq:refEquation28}), i.e., when an ion crosses the surface of a sphere of the radius $R_b$ then its velocity ${\mathbf{v}}_{i}$ is transformed as
\begin{equation}
{{\mathbf{v}}'_i} = {{\mathbf{v}}_i} - \frac{{2{{\mathbf{r}}_i}{{\mathbf{v}}_i}}}{{r_i^2}}{{\mathbf{r}}_i}.
\end{equation}
Due to the finiteness of the time integration step, the ion velocity can be transformed only at the point where $r_i > R_b$, i.e., the ion density distribution $f_c(r)$ is never step-like but its vanishing at $r > R_b$ is smooth. Were we set $R_b = R$ then some artificial charged layer would form beyond the surface of compensating background, which violates the charge neutrality inside the BOCP. This effect cannot be eliminated completely but it can be minimized by slightly reducing $R_b$ so that it coincides with the ion equimolar radius. Since the average distance at which an ion propagates during the time step $\tau$ is $\tau /\sqrt {2\pi \Gamma }$, the shift $R_b = R-\tau /\sqrt {8\pi \Gamma }$ minimizes the boundary effect. Note that increasing or decreasing the shift by several times does not significantly change the results but the BOCP size beginning from which the size dependence $u(\Gamma)$ corresponds to Eq.\ (\ref{seq:refEquation18}) increases noticeably thus indicating the increase of the boundary effect. Additionally, at each time step and for each ion, the components of ion velocity were scaled such that its modulus was set to $\sqrt{2/\Gamma}$ if it exceeded $\sqrt{25/\Gamma}$.

\begin{table}
\caption{\label{t1}Parameters of the MD simulation for BOCP at different $\Gamma$'s. Given are the time integration step $\tau$, the range of ion numbers $N$, and the range of respective time step numbers $N_{\tau}$ (sample sizes are given in parenthesis).}
\begin{ruledtabular}
\begin{tabular}{cccc}
 ${\Gamma}$ & ${\tau}$ & Range of $N$ & Range of $N_\tau\times 10^{-4}$ \\ \hline
0.03 & 0.001 & 5000--50000 & 570(286)--6.67(3.33) \\
0.1 & 0.002 & 2500--50000 & 1800(900)--6.67(3.33) \\
0.3 & 0.01 & 2500--50000 & 1800(900)--6.67(3.33) \\
1 & 0.01 & 2500--50000 & 1800(900)--6.67(3.33) \\
3 & 0.01 & 2500--50000 & 1800(900)--6.67(3.33) \\
10 & 0.01 & 2500--50000 & 1800(900)--6.67(3.33) \\
17 & 0.01 & 2500--50000 & 1800(900)--6.67(3.33) \\
30 & 0.01 & 2500--50000 & 1800(900)--6.67(3.33) \\
100 & 0.01 & 2500--50000 & 1800(900)--6.67(3.33) \\
174 & 0.05 & 2500--50000 & 1800(900)--6.67(3.33) \\
176 & 0.05 & 2500--50000 & 1800(900)--6.67(3.33) \\
178 & 0.05 & 2500--50000 & 3600(900)--13.4(3.33) \\
182 & 0.05 & 2500--50000 & 3600(900)--13.4(3.33) \\
190 & 0.05 & 2500--50000 & 3600(900)--13.4(3.33) \\
200 & 0.05 & 2500--10000 & 3600(900)--304(76.0) \\
210 & 0.05 & 2500--10000 & 3600(900)--304(76.0) \\
300 & 0.05 & 2500--50000 & 5400(900)--20.0(3.33) \\
650 & 0.1 & 2500--21200 & 5400(900)--109(18.2) \\
1000 & 0.1 & 2500--30000 & 5400(900)--54.9(9.2) \\
\end{tabular}
\end{ruledtabular}
\end{table}
The choice of time integration step depends on $\Gamma$, and $\tau(\Gamma)$ is an increasing function due to slowdown of the thermal motion. If $\tau$ is too small then too large number of time steps $N_{\tau}$ that exceeds the computational resources is required to equilibrate the system. For too large $\tau$, the accuracy of the used numerical integration scheme becomes insufficient, which leads to noticeable errors in the parameters of interest. Thus, an optimum $\tau$ was determined for each $\Gamma$ from the condition that the determined parameters are almost insensitive to variation of $\tau$ by several times (Table~\ref{t1}).

MD simulation of BOCP was performed for the set of BOCP sizes $N=2500$, 3540, 5000, 7070, 10000, 14000, 21200, 30000, and 50000 at different $\Gamma$. Due to the limitation of computational resources, the maximum $N$ is defined by the minimum $N_{\tau}$, for which variation of the determined quantity is much less than its estimated error. Thus, in some cases, simulations were performed in a limited range of $N$. The ranges of $N$ and $N_{\tau}$ estimated in such a way for different $\Gamma${}'s are also listed in Table~\ref{t1}. For the sizes within the range limits, $N_{\tau}\propto N^{-2}$. The upper bound of $N$ at $\Gamma=200$ and 210 is a consequence of higher fluctuations in the system, where a transitional regime from fluid to solid (fluid core and solid shells) takes place. At $\Gamma=650$ and 1000, the BOCP equilibration time increases sharply due to the decrease in the ion self-diffusion coefficient for a solid system. For $\Gamma=0.03$, the lower bound of $N$ is due to the fact that the size dependence $u(\Gamma)$ does not correspond to Eq.\ (\ref{seq:refEquation18}) at $N<3540$.

First, the positions of $N$ ions were initialized using the homogeneous random spatial distribution in a sphere of the radius $R_b$. Then Eqs.\ (\ref{seq:refEquation28}) were integrated numerically using the Verlet algorithm. After the relaxation of the system to equilibrium over $N_{\tau}/2$ time steps, data were collected every 100 time steps. For the calculation of $f_p(r)$ (\ref{seq:refEquation6}), an ion closest to the BOCP center was selected at the same time interval. The data for different quantities were averaged, and the error was estimated specifically for each quantity (Sec.~\ref{bkm:RefNumPara9096183868828}). Eventually, the obtained data were used for the determination of the OCP properties in the thermodynamic limit.

\subsection{Simulation results}
\label{bkm:RefNumPara9096183868828}Figure~\ref{f1} shows typical ion density distributions and the radial distribution functions obtained in our simulations. Largely, in the BOCP core region, the radial distribution functions are almost identical to those obtained for the OCP in various MC simulations starting from \cite{Brush_1966} (Fig.~\ref{f1}b). Namely, a dip in $f_p(r)$ at small distances for $\Gamma=0.1$ is indicative of repulsion; the radial distribution function exhibits two maxima indicating a smooth transfer from the gaslike to fluid state at $\Gamma=30$; and several maxima emerge at $\Gamma=200$, which is somewhat below the melting point for the size $N=5000$. The maxima at larger $r$ are due to the shells formed closer to the BOCP boundary (two shells can be seen already at $\Gamma=30$). Note that melting of shells is observed at $\Gamma=60\mbox{--}90$ well below the melting point of the central part of the Coulomb clusters \cite{Shpil_ko_2023}. In Fig.~\ref{f1}a, the ion density distribution $f_c(r)$ demonstrates qualitatively the same regularities: in most of the BOCP volume, the ions are homogeneously distributed at $\Gamma=0.1$ and 30, although two shells are visible for $\Gamma=30$. However, the curve for $\Gamma=200$ is more indicative of the shell structure than in Fig.~\ref{f1}b because in the latter, departure of the ion, for which $f_p(r)$ is determined, from the BOCP center smooths the peaks of shells considerably. Instead, as follows from Fig.~\ref{f1}a, shells fill a significant part of the BOCP volume at high $\Gamma$. As is seen in Fig.~\ref{f1}c, both $f_p(r)$ and $f_c(r)$ exhibit behavior with irregular tooth-shaped peaks typical for a solid. If we model the BOCP by a quasi-uniform core surrounded by shells then the radius of mesoscopic uniformity can be estimated as $N_\mr{cr}^{1/3}$, where $N_\mr{cr}$ is the number of ions forming a 3D crystallized core (not belonging to shells). As follows from the results of \cite{Zhukhovitskii_2024}, for $N=5000$, $N_\mr{cr}^{1/3}\simeq 6.88$ at $\Gamma=210$ and $13.4$ at $500$, while the cluster radius is $R=17.1$ (see also Fig.~4 in \cite{Zhukhovitskii_2024}).

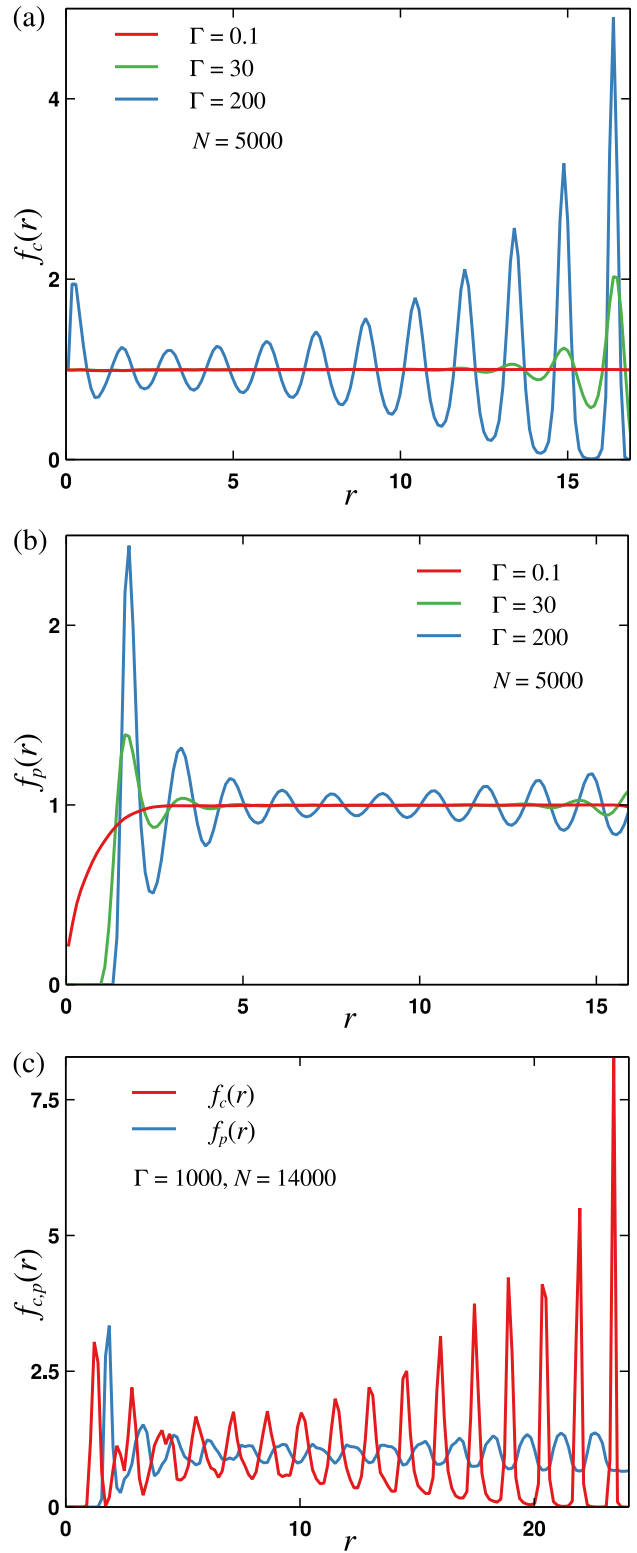
\begin{myfigure}{0.93}{1}{Ion distribution functions: (a) the ion density distribution $f_c(r)$ (\ref{seq:refEquation7}) and (b) the radial distribution function $f_p(r)$ (\ref{seq:refEquation6}) of BOCP at different $\Gamma$; (a) and (b) $\Gamma=0.1$, 30, and 200 (red, green, and blue lines, respectively), $N=5000$; (c) $f_c(r)$ (red line) and $f_p(r)$ (blue line) at $\Gamma=1000$ and $N=14000$. The BOCP radii are (a) 17.1, (b) 17.1, and (c) 24.1.}

\end{myfigure}

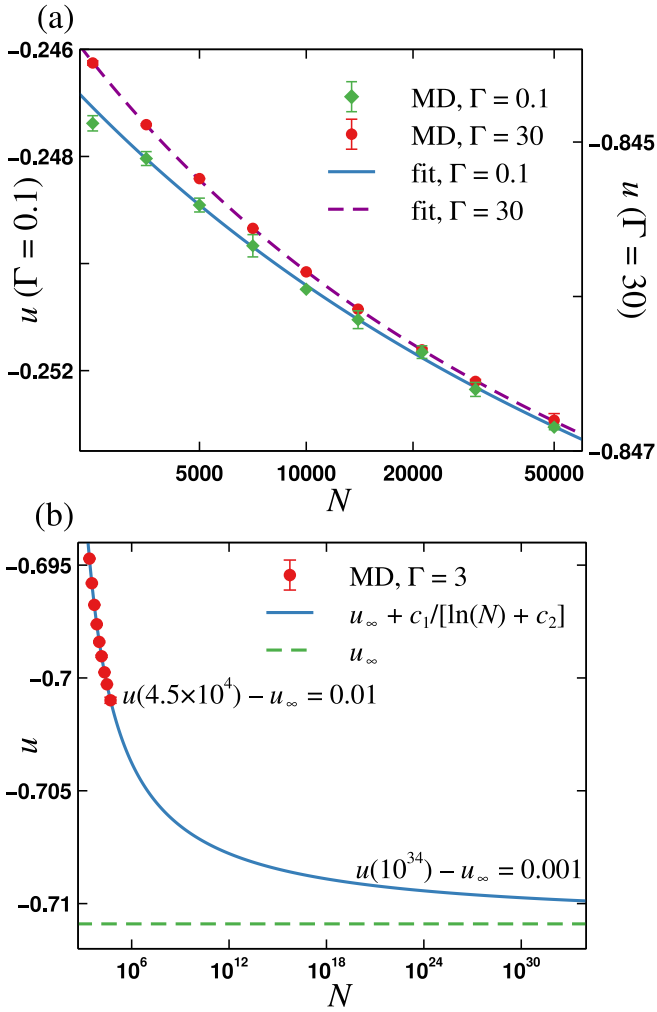
\begin{myfigure}{0.93}{2}{Size dependence of the BOCP total electrostatic energy from MD simulation (dots) and its fit by Eq.\ (\ref{seq:refEquation18}) (lines): (a) diamonds and solid lines correspond to $\Gamma=0.1$ and circles and dashed line, to $\Gamma=30$; (b) circles indicate MD data for $\Gamma=3$, solid line is their fit, and dashed line marks the thermodynamic limit $u_\infty(3)$.}

\end{myfigure}

The size dependence of $u(\Gamma,N)$ obtained from MD simulation is illustrated by Fig.~\ref{f2}a, which testifies the accuracy of its approximation by the function (\ref{seq:refEquation18}). Indeed, the actual deviation of this approximation from the obtained MD results for all BOCP sizes, e.g., for $\Gamma=30$, varies from $4\times10^{-8}$ to $8\times10^{-6}$. The accuracy of MD results determined by a standard statistical method depending on the sample size is of the same order of magnitude and in principle it can be further decreased by the increase in $N_\tau$. However, the difference in $u(\Gamma,N)$ obtained for two similar runs initialized with different random seeds is orders of magnitude greater, which means that a real simulation error is of physical nature. Thus, this difference may be due to the low-frequency non-vanishing wing of the phonon spectrum when the inverse phonon frequency is of the same order as $\tau N_\tau$, or it may arise from different structures of the polycrystal formed at $\Gamma \gg 1$. Thus, for each $\Gamma$ and $N$, the simulation was repeated twice; $u(\Gamma,N)$ was then estimated as the average of these two results while the accuracy was set to the modulus of their difference. For $\Gamma=30$, such accuracy is not worse than $4\times 10^{-5}$, which is noticeably greater than the accuracy of data approximation. The same procedure was applied for all determined size-dependent quantities. It is this accuracy that is shown by error bars in Figs.~\ref{f2}, \ref{f4}, and \ref{f6}--\ref{f8}. Note that in most figures, the error bar width is less than the dot marker size.

Albeit the function (\ref{seq:refEquation18}) approximates $u(\Gamma,N)$ with a high accuracy, the size dependence is very weak at great $N$. This makes extrapolation of $u(\Gamma,N)$ to the thermodynamic limit problematic. In other terms, it is necessary to estimate the accuracy of $u_\infty(\Gamma)$ obtained from such extrapolation. To this end, from the set of data for the same $\Gamma$, we selected at random two triplets of unmatched BOCP sizes $N$ and fitted (\ref{seq:refEquation18}) to these triplets. Modulus of the half-difference between two values of $u_\infty(\Gamma)$ obtained in such a way is associated with the accuracy of its determination in Figs.~\ref{f3} and \ref{f5}. As is seen in Table~\ref{t2}, the overall relative accuracy of extrapolated quantities is of the order of 0.1\%, which is greater than the errors for individual values but still reasonably good. Figure~\ref{f2}b illustrates an extremely slow convergence of $u(\Gamma,N)$ to $u_\infty(\Gamma)$. As is seen, the accuracy of 1\% is reached already for $N=45000$, which is close to the size used in this work, while the overall approximation accuracy of 0.1\% is attained only for $N=10^{34}$, i.e., for a large body.

\begin{table*}
\caption{\label{t2}BOCP parameters in the thermodynamic limit. Listed are the total potential energy $u_{\infty}$ and its fit $u_\mr{fit}$ (\ref{seq:refEquation37}), (\ref{seq:refEquation38}); the excess energy of interionic interaction $u_{p\mr{ex}}$, the excess energy of interaction between the ions and background $u_{b\mr{ex}}$ (listed are the results for $N=21200$), and the shifted ion compressibility factor $Z_{i}-1$ calculated from the excess energies (\ref{seq:refEquation15}) and virial (\ref{seq:refEquation23}).}
\begin{ruledtabular}
\begin{tabular}{ccccccc}
 ${\Gamma}$ & $u_{\infty}$ & $u_\mr{fit}$ & $u_{p\mr{ex}}$ & $2u_{b\mr{ex}}$ & $Z_{i}-1$ (energy) & $Z_{i}-1$ (virial) \\ \hline
0.03 & $ -0.1466\pm 0.0027$ & -- & $ -0.1402\pm 0.0093$ & $ -0.0059\pm 0.018$ & $-0.00157\pm 0.0002$ & $ -0.00143\pm 0.001$ \\
0.1 & $ -0.2659\pm 0.0025$ & $-0.2659$ & $ -0.2391\pm 0.0151$ & $ -0.0261\pm 0.030$ & $ -0.00779\pm 0.0002$ & $ -0.00798\pm 0.001$ \\
0.3 & $ -0.4050\pm 0.0050$ & $-0.4050$ & $ -0.3726\pm 0.0010$ & $ -0.0409\pm 0.002$ & $ -0.0332\pm 0.0001$ & $ -0.0343\pm 0.0005$ \\
1 & $ -0.5785\pm 0.0016$ & $-0.5784$ & $ -0.5095\pm 0.0002$ & $ -0.1139\pm 0.0004$ & $ -0.1317\pm 0.0004$ & $ -0.1320\pm 0.0008$ \\
3 & $ -0.7109\pm 0.0005$ & $-0.7111$ & $ -0.5941\pm 0.00002$ & $ -0.2113\pm 0.00002$ & $ -0.3834\pm 0.0004$ & $ -0.3853\pm 0.0011$ \\
10 & $ -0.8044\pm 0.0012$ & $-0.8041$ & $ -0.6132\pm 0.00008$ & $ -0.3665\pm 0.0002$ & $ -0.8241\pm 0.0008$ & $ -0.8242\pm 0.0018$ \\
17 & $-0.8304\pm 0.0001$ & $-0.8302$ & $-0.6056\pm 0.00009$ & $-0.4375\pm 0.0002$ & $-0.9531\pm 0.0004$ & $-0.9443\pm 0.0057$ \\
30 & $ -0.8506\pm 0.0002$ & $-0.8507$ & $ -0.5975\pm 0.00003$ & $ -0.4978\pm 0.00004$ & $ -0.9954\pm 0.0006$ & $ -0.9988\pm 0.0026$ \\
100 & $ -0.8762\pm 0.0010$ & $-0.8765$ & $ -0.5930\pm 0.00004$ & $ -0.5630\pm 0.00008$ & $ -0.9976\pm 0.0010$ & $ -0.9851\pm 0.0070$ \\
174 & $ -0.8834\pm 0.0001$ & $-0.8831$ & $ -0.5939\pm 0.00002$ & $ -0.5766\pm 0.00002$ & $ -0.9984\pm 0.0013$ & $ -1.0027\pm 0.0039$ \\
176 & $ -0.8835\pm 0.0001$ & $-0.8833$ & $ -0.5939\pm 0.00001$ & $ -0.5768\pm 0.00001$ & $ -0.9997\pm 0.0007$ & $ -0.9984\pm 0.0020$ \\
178 & $ -0.8841\pm 0.0002$ & $ -0.8866 $ & $ -0.5939\pm 0.020$ & $ -0.5771\pm 0.051$ & $ -0.9982\pm 0.0012$ & $ -0.9951\pm 0.0038$ \\
182 & $ -0.8850\pm 0.0002$ & $ -0.8869 $ & --- &  ---  & $ -0.9973\pm 0.0015$ & $ -0.9961\pm 0.0018$ \\
190 & $ -0.8881\pm 0.0007$ & $ -0.8874 $ & --- & --- & $ -0.9983\pm 0.0016$ & $ -0.9955\pm 0.0022$ \\
200 & $ -0.8913\pm 0.0045$ & $ -0.8879 $ & --- & --- & $ -0.9992\pm 0.0011$ & $ -0.9940\pm 0.0036$ \\
210 & $ -0.8950\pm 0.0025$ & $ -0.8883 $ & --- & --- & $ -0.9984\pm 0.0005$ & $ -0.9959\pm 0.0016$ \\
300 & $ -0.8909\pm 0.0009$ & $ -0.8909 $ & $ -0.5958\pm 0.00001$ & $ -0.5859\pm 0.00002$ & $ -0.9992\pm 0.0012$ & $ -0.9994\pm 0.0094$ \\
650 & $ -0.8939\pm 0.0012$ & $ -0.8939 $ & $ -0.5958\pm 0.00002$ & $ -0.5912\pm 0.00002$ & $ -0.9947\pm 0.0022$ & $ -0.9860\pm 0.0068$ \\
1000 & $ -0.8948\pm 0.0002$ & $ -0.8948 $ & $ -0.5958\pm 0.00001$ & $ -0.5928\pm 0.00005$ & $ -0.9949\pm 0.0029$ & $ -1.0134\pm 0.0126$ \\
\end{tabular}
\end{ruledtabular}
\end{table*}

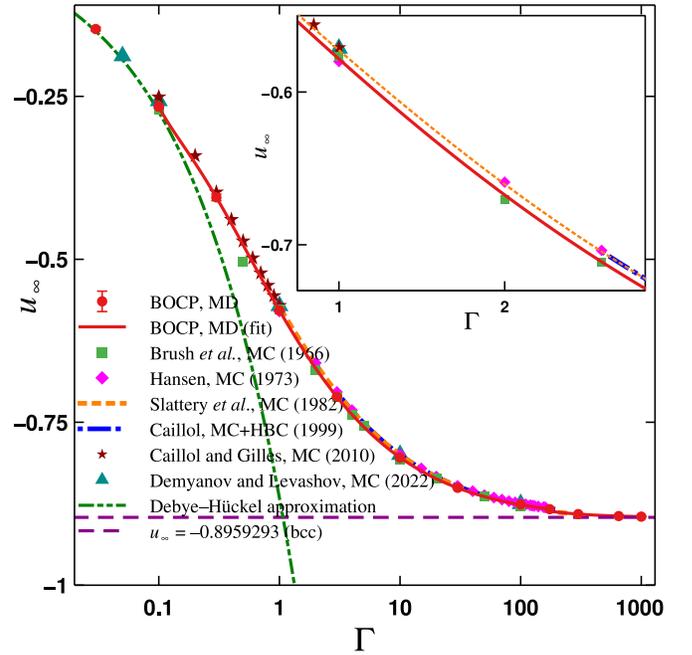
\begin{myfigure}{0.978}{3}{Thermodynamic limit of the BOCP electrostatic energy compared with the results of MC simulations for the OCP with PBC. Shown are the MD results of this work (circles) and their approximation by the functions (\ref{seq:refEquation37}) and (\ref{seq:refEquation38}) (solid line), MC results by Brush \textit{et al.} \cite{Brush_1966} (squares), Hansen \cite{Hansen_1973} (diamonds), Caillol and Gilles \cite{Caillol_2010} (stars), Demyanov and Levashov \cite{Demyanov_2022} (triangles). Approximation of the results available in studies by Slattery \textit{et al.} \cite{Slattery_1982} and Caillol (MC within hyperspherical boundary conditions) \cite{Caillol_1999_3} are shown by dotted and dashed{}-dotted lines, respectively. Dashed{}-dotted{}-dotted line indicates the Debye--Hückel approximation and dashed line, the Madelung constant for bcc lattice. Inset shows a magnified fragment of this figure.}

\end{myfigure}

The total electrostatic energy $u_\infty(\Gamma)$ determined from MD simulation is listed in Table~\ref{t2}. These data were fitted by the function taken from \cite{Caillol_1999_3}
\begin{equation}\label{seq:refEquation37}
\Gamma {u_{\mr{fit}}} = A + B\Gamma  + C{\Gamma ^s} + D{\Gamma ^{ - s}},\quad 0.1 \le \Gamma  \le 174
\end{equation}
with $A=-0.830421$, $B=-0.897496$, $C=0.950025$, $D=0.19952$, and $s=0.239635$, and by the theoretical dependence (cf.\ (\ref{seq:refEquation24}) and (\ref{seq:refEquation25}))
\begin{equation}\label{seq:refEquation38}
{u_{\mr{fit}}} = {\beta _0} + \frac{{{\beta _1}}}{\Gamma } + \frac{{{\beta _2}}}{{{\Gamma ^2}}},\quad 174 < \Gamma  \le 1000
\end{equation}
with $\beta_0=-0.896383$, $\beta_1=1.5311$, and $\beta_2=35.0926$. Note that at $\Gamma\le176$, $\left| {{u_\infty } - {u_{\mr{fit}}}} \right| \sim {10^{ - 4}}$, which is much less than the error in $u_\infty$ determination.

The parameters in Eq.\ (\ref{seq:refEquation38}) $\beta_0$ and $\beta_1$ obtained from the fitting procedure prove to be close to those calculated in the harmonic lattice approximation: $\beta_0\approx u_0$ with the same accuracy of $\sim10^{-3}$ as in Table~\ref{t2} and $\beta_1\approx 3/2$. The first anharmonicity correction coefficient turns out to be of the same order of magnitude as calculated in study by Chugunov and Baiko \cite{Chugunov_2005} for the classical OCP, where $\beta_2\approx10.58$. Apparently, the accuracy of $\beta_2$ determination could have been improved if more points were calculated in the interval $174 < \Gamma \le 1000$ (at present, only three were fitted), and it might bring our result closer to \cite{Chugunov_2005}. Even better accuracy in $\beta_0$ is attained if we fit solely this parameter taking the theoretical values $\beta_1=1.5$ and $\beta_2=10.58$. Here, $\beta_0=-0.896204$, which is less than $u_0$ by only $2.8\times10^{-4}$ . One can conclude that the energy of a cold system tends to $u_0$, which is in agreement with the result of work by Hasse \cite{Hasse_2003}.

Determined energies are compared with the results of different studies in Fig.~\ref{f3}. As is seen, they are in a good agreement with early data by Brush \textit{et al.} \cite{Brush_1966} and Hansen \cite{Hansen_1973}, however, dispersion of those data are much greater than the accuracy attained in this work. Note that here, our error bars are smaller than the dot marker sizes. The results obtained later by Slattery \textit{et al. }\cite{Slattery_1982} are in perfect agreement with modern data from the works by Demyanov and Levashov \cite{Demyanov_2022}, and by Caillol \cite{Caillol_1999_3} and Caillol and Gilles \cite{Caillol_2010} for $\Gamma\ge1$, while at $\Gamma<1$, the data from \cite{Caillol_2010} are situated higher than any other data. At $\Gamma\le0.1$, our results scarcely deviate from the Debye--Hückel approximation, while the data from \cite{Caillol_2010} are noticeably higher. At $\Gamma>174$, our data reach asymptotically the bcc energy $u_0$, and in this region, they are in a very good agreement with \cite{Slattery_1982} (Fig.~\ref{f5}). At the same time, for $0.1<\Gamma<174$, our energies are appreciably lower than other data, which can be clearly seen in the inset in Fig.~\ref{f3}. Here, a typical difference is about 0.5\%, which is much higher than the declared accuracy of all results. Apparently, this difference may arise from the limitation on system microscopic states imposed by the PBC, which can be qualitatively treated as strong correlations between the particles in a supercell and its images. Since the characteristic decay length of the Ewald potential is of the order of the simulation cell length, the resulting long-range interaction may contribute noticeably to the total system energy. In contrast, BOCP does not have such limitation, and this leads the system to a more favorable state. Nevertheless, at $\Gamma \to \infty$, the system occupies a single state, which results in closeness of our data and \cite{Slattery_1982}.

Based on Eqs.\ (\ref{seq:refEquation37}) and (\ref{seq:refEquation38}), we can approximate the excess internal energy in a wide range of $\Gamma$ in the form
\begin{equation}
{\varepsilon _{\mr{ex}}}(\Gamma ) = \left\{ {\begin{array}{*{20}{c}}
{{u_{\mr{fit}} }(\Gamma) - {u_0},}&{0.1 \le \Gamma  \le {\Gamma _\mr{m}},}\\
{ \frac{3}{{2\Gamma }} + \frac{{{\beta _2}}}{{{\Gamma ^2}}},}&{\Gamma  > {\Gamma _\mr{m}},}
\end{array}} \right.
\end{equation}
and the excess Helmholtz free energy, as
\begin{equation}
\eqal{
{f_{\mr{ex}}}(\Gamma ) & = \int {\frac{{{\varepsilon _{\mr{ex}}}(\Gamma )}}{\Gamma }} {\mkern 1mu} {\kern 1pt} d\Gamma\\ 
& = \left\{ {\begin{array}{*{20}{c}}
{{\omega _1}(\Gamma ) + \Delta {\omega _1},}&{0.1 \le \Gamma  \le {\Gamma _\mr{m}},}\\
{{\omega _2}(\Gamma ),}&{\Gamma  > {\Gamma _\mr{m}},}
\end{array}} \right.
}
\end{equation}
where
\begin{equation}
{\omega _1}(\Gamma ) = (B - {u_0})\ln \Gamma  - \frac{A}{\Gamma } + \frac{{C{\Gamma ^{s - 1}}}}{{s - 1}} - \frac{{D{\Gamma ^{ - s - 1}}}}{{s + 1}},
\end{equation}
${\omega _2}(\Gamma ) =  - {\beta _1}/\Gamma  - {\beta _2}/2{\Gamma ^2}$, $\Delta {\omega _1} = {\omega _2}({\Gamma _\mr{m}}) - {\omega _1}({\Gamma _\mr{m}})$, and we take into account that $f_{\mr ex}(\Gamma )$ is a continuous function at $\Gamma=\Gamma_{\mr m}$.

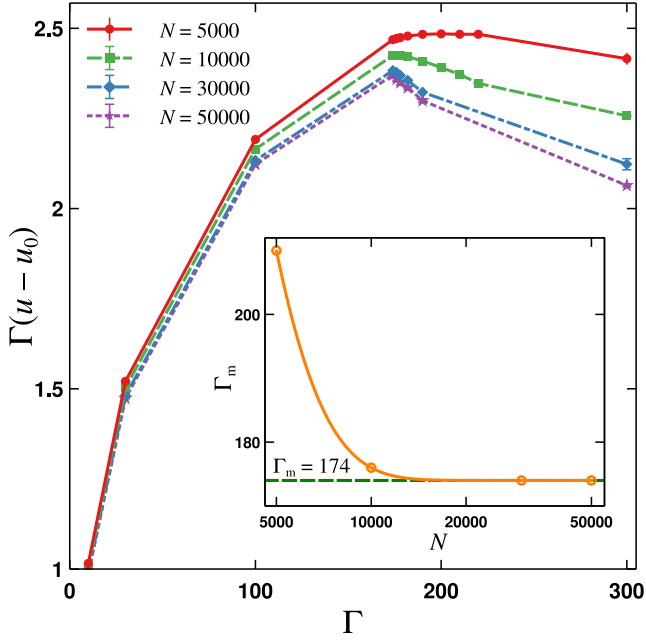
\begin{myfigure}{0.978}{4}{Thermal fraction of the BOCP electrostatic energy as a function of $\Gamma$ for the BOCP size $N=5000$ (solid line), 10000 (dashed line), 30000 (dashed{}-dotted line), and 50000 (dotted line). Inset shows the size-dependent Coulomb coupling parameter $\Gamma_{\mr{m}}(N)$ (open circles); solid line indicates the curve fit and dashed line, the thermodynamic limit $\Gamma_{\mr{m}}(\infty)$.}

\end{myfigure}

A good parameter for investigation of melting is the thermal fraction of the Coulomb energy $\Gamma[u(\Gamma, N) - u_{0}]$ \cite{Slattery_1982}. It is well known that, in contrast to the bulk system, phase transitions in finite-size systems are very diffuse, both in temperature and phase composition. As can be seen in Fig.~\ref{f4}, the BOCP is no exception. In this case, melting can be associated with the Coulomb coupling parameter $\Gamma_{\mr{m}}$ corresponding to the maximum of $\Gamma_{\mr{m}}[u(\Gamma_{\mr{m}}, N) - u_{0}]$. This maximum is shifted toward lower values with the increase in $N$. The inset shows a fit of $\Gamma_{\mr{m}}$ to the exponential decay function $\Gamma_{\mr{m}}(N)=174.0+647.995\exp(-N/1729.886)$, from which the estimate for the bulk melting parameter is $\Gamma_{\mr{m}}(\infty)=174\pm2$. The error of 2 corresponds to the error of maximum resolution for the size-dependent function $\Gamma_{\mr{m}}[u(\Gamma_{\mr{m}}, N) - u_{0}]$ due to the error in $u(\Gamma,N)$. Obtained value correlates with that obtained in \cite{Slattery_1982} ($\Gamma_{\mr m}=178$) and coincides with the results by Dubin and O'Neil \cite{Dubin_1999} and Khrapak and Khrapak \cite{Khrapak_2016}. The absence of an appreciable metastable region in the neighborhood of $\Gamma_{\mr{m}}$ is noteworthy. It means that the fluid--solid transition is very fast, at least, its characteristic time is less than $\tau N_\tau$. This can be connected with the solid-like shells near the BOCP boundary, which initiate crystallization propagating to the BOCP center.

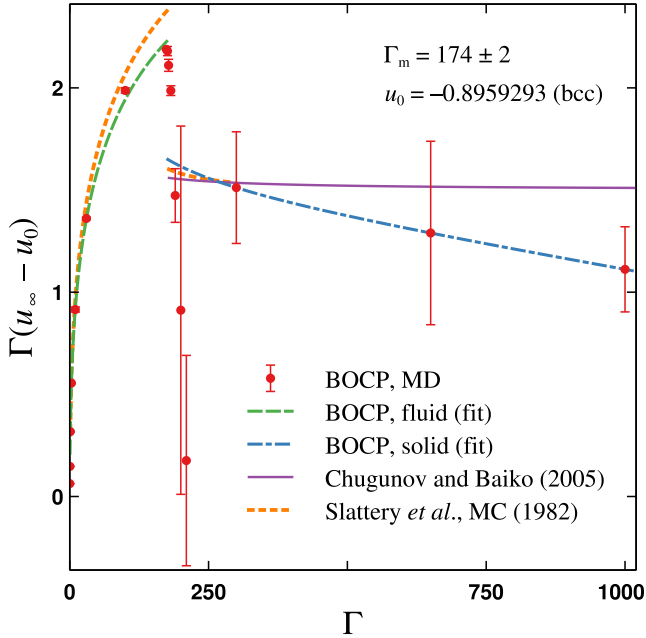
\begin{myfigure}{0.978}{5}{Thermal fraction of the electrostatic energy in the thermodynamic limit as a function of $\Gamma$. This work: MD results (dots) and their approximation by the function (\ref{seq:refEquation37}) (dashed line) and (\ref{seq:refEquation38}) with three fitting parameters (dashed{}-dotted line). Solid line indicates calculations based on the harmonic lattice model supplemented by anharmonicity correction \cite{Chugunov_2005} and dotted line presents MC results \cite{Slattery_1982}.}

\end{myfigure}

Another way to estimate $\Gamma_{\mr{m}}$ is based on the parameter $\Gamma(u_{\infty} - u_{0})$ for the bulk fluid and solid (Fig.~\ref{f5}). Here, in contrast to Fig.~\ref{f4}, the dependence of this parameter on $\Gamma$ forms a cliff at $\Gamma_{\mr{m}}=174\pm2$, where again, the estimation for accuracy follows from the error in $u_{\infty}(\Gamma)$. Note that at $190\le\Gamma\le210$, the dots are situated well below the data for $\Gamma\ge300$. This is a consequence of a poor applicability of the principal size dependence (\ref{seq:refEquation18}) for $N>10000$, owing to which the data in Table~\ref{t2} were obtained only for the smaller sizes. Very large errors of $u_\infty$ calculations in this range of $\Gamma$ are indicative of enormous fluctuations of the BOCP structure at $\Gamma\approx\Gamma_{\mr m}$. Here, $N_{\tau}$ limited by the computational resources and listed in Table~\ref{t1} seems to be insufficient to reach equilibration for such sizes. However, for $\Gamma\le178$ and $\Gamma\ge300$, the systems are sufficiently close to equilibrium. One can testify that at high $\Gamma$, our data agree qualitatively with the analytical results by Chugunov and Baiko \cite{Chugunov_2005}. Note that, as for a finite size BOCP, we have found no metastable region. In addition, it is worth mentioning that the excess free energy at the melting point (\ref{seq:refEquation27}) amounts to $0.0092$, which proves to be an order of magnitude less than what follows from the estimate of free energy in \cite{Slattery_1982}.

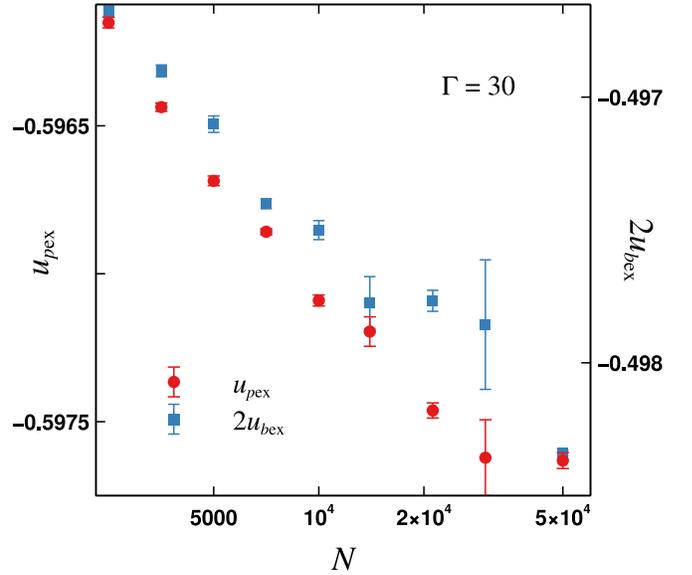
\begin{myfigure}{0.978}{6}{Size dependences of the excess interatomic electrostatic energy $u_{p{\mr ex}}$ (circles) and the excess ion--background electrostatic energy $u_{b{\mr ex}}$ (triangles) at $\Gamma=30$.}

\end{myfigure}

Now we turn to the determination of the excess interatomic and ion--background electrostatic energy (Table~\ref{t2} and Fig.~\ref{f6}). The size dependences of excess energies $u_{p\mr ex}(\Gamma,N)$ and $u_{b\mr ex}(\Gamma,N)$ (\ref{seq:refEquation4}) determined in our simulation reveal an overall trend to decrease with the increase in $N$. However, the decrease is so small that we failed to assign any functional dependence to it in the investigated range of system parameters. Rigorously speaking, we are not able to extrapolate the results to the thermodynamic limit due to an extremely narrow abscissa range in Fig.~\ref{f6} and hence, confine ourselves to its estimate by taking these quantities for $N=21200$. Here, the corresponding errors were calculated in the same way as for $u(\Gamma,N)$. The resulting dependences $u_{p{\mr ex}}(\Gamma)$ and $u_{b{\mr ex}}(\Gamma)$ are shown in Fig.~\ref{f7} and Table~\ref{t2}. Since at $182\le\Gamma\le210$, BOCP of the sizes $N>10000$ were not simulated, the corresponding $u_{p{\mr ex}}$ and $u_{b{\mr ex}}$ are not listed in the table. Behavior of these dependences is in agreement with the asymptotic analytical predictions (\ref{seq:refEquation11}) and (\ref{seq:refEquation17}), which is indicative of the fact that Fig.~\ref{f7} represents a satisfactory estimate for the investigated quantities in the thermodynamic limit. One can see that as $\Gamma$ is decreased, $u_{b{\mr ex}}(\Gamma)$ vanishes more rapidly than $u_{p{\mr ex}}(\Gamma)$, which can be explained by the difference in $f_p(r)$ (\ref{seq:refEquation6}) and $f_c(r)$ (\ref{seq:refEquation7}). Indeed, at $\Gamma=0.1$, $f_c\simeq 1$ (Fig.~\ref{f1}a), which vanishes the integral (\ref{seq:refEquation9}), whereas $f_p$ is noticeably different from zero at short distances (Fig.~\ref{f1}b) resulting in nonzero $u_{p{\mr ex}}(\Gamma)$ (\ref{seq:refEquation8}). Since (\ref{seq:refEquation8}) and (\ref{seq:refEquation9}) were written in the approximation of pair correlations for a homogeneous medium, it is clear that such approximation is ineffective for $u_{b{\mr ex}}(\Gamma)$, whose analytical calculation would require the inclusion of higher-order correlations. In this connection, a fast decrease of $u_{b{\mr ex}}(\Gamma)$ at $\Gamma>0.3$ is a consequence of inadequacy of the Debye--Hückel approximation in this range of $\Gamma$. Notably, it follows from the recent analysis \cite{Khrapak_2025} of the reduced coefficients of self-diffusion, shear viscosity, and thermal conductivity of the OCP based on the Debye--Hückel plus hole approach \cite{Nordholm_1984} that the binary collision approaches break down at $\Gamma>0.3$, when OCP is no more a weakly coupled system.

At high $\Gamma$, $u_{p{\mr ex}}(\Gamma)$ reaches a solid state value already at $\Gamma=30$ while $u_{b{\mr ex}}(\Gamma)$, only at $\Gamma=650$. Interestingly, $u_{p{\mr ex}}(\Gamma)$ has a shallow minimum at $\Gamma\approx10$, which coincides with a gas-to-liquid dynamical crossover in the OCP discussed in works \cite{Huang_2023} and \cite{Yu_2024}, so this minimum can serve as yet another definition of the Frenkel line in OCP.

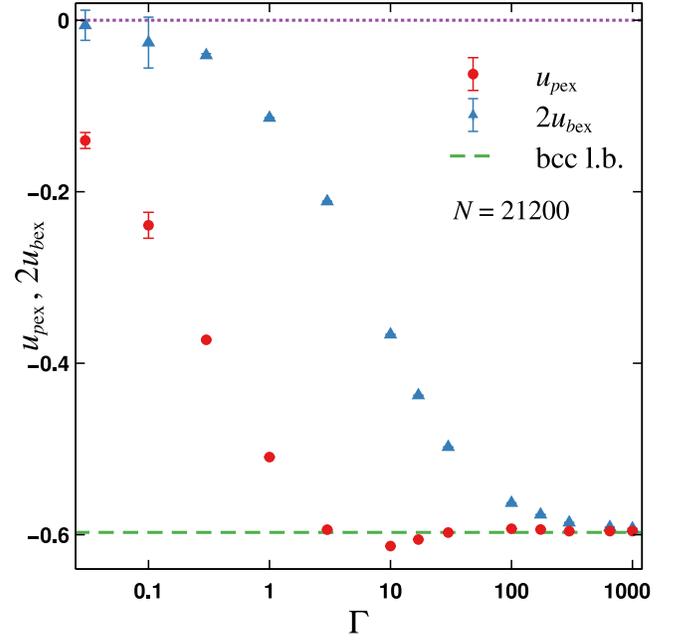
\begin{myfigure}{0.978}{7}{Excess interatomic electrostatic energy (circles) and the excess ion--background electrostatic energy (triangles) as functions of $\Gamma$. Dashed line indicates the bcc lower bound for these quantities; $N=21200$.}

\end{myfigure}

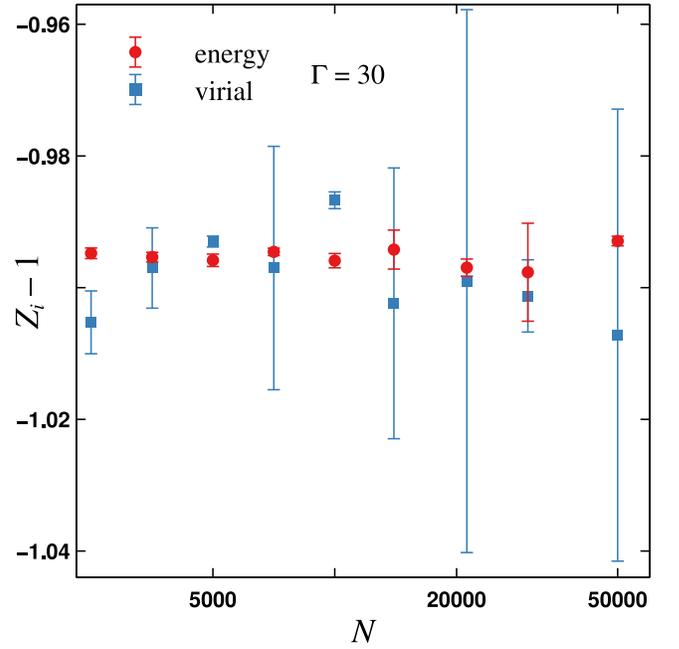
\begin{myfigure}{0.978}{8}{Size dependence of the ionic compressibility factor $Z_i$ determined from MD simulation using Eqs.\ (\ref{seq:refEquation15}) (circles) and (\ref{seq:refEquation23}) (squares) at $\Gamma=30$.}

\end{myfigure}

Determination of the energies $u_{p\mr ex}(\Gamma,N)$ and $u_{b\mr ex}(\Gamma,N)$ allows one to obtain the ionic compressibility factor (\ref{seq:refEquation15}). In Fig.~\ref{f8}, it is compared with the same quantity estimated from the virial (\ref{seq:refEquation23}). It is noteworthy that (i) estimates from the energies are in a good agreement with those from the virial, however, the latter have much larger scatter and (ii) within the accuracy attained in our simulation, $Z_i$ seems to be size-independent. Such peculiarities hold true irrespective of $\Gamma$. Thus, we take the average of all $Z_i$ available for different $N$ at the same $\Gamma$ as an estimate of its thermodynamic limit, the error being calculated by application of a standard statistical method to the sample of size up to nine, by the number of BOCP sizes used. The resulting dependence $Z_i(\Gamma)$ is shown in Fig.~\ref{f9} and listed in Table~\ref{t2}. As is seen, the results from energy and virial are hardly distinguishable, except for the highest $\Gamma$, which validates the overall correctness of $Z_i$ determination. At $\Gamma<0.3$, the obtained $Z_i$ almost coincides with the ionic compressibility factor $Z=1 - {\Gamma ^{3/2}}/2\sqrt 3$ that results from the Debye--Hückel approximation, as it must. In the interval $0.1\le\Gamma\le27.72$, $Z_i$ can be approximated by the function
\begin{equation}\label{seq:refEquation42}
{Z_i}(\Gamma ) = {c_1} + {c_2}\tanh \left( {\frac{{{c_3} - \ln \Gamma }}{{{c_4}}}} \right)
\end{equation}
with $c_1=0.461376$, $c_2=0.526010$, $c_3= 1.483011$, and $c_4=1.350840$. Note that at $\Gamma>27.72$, $Z_i \simeq 0$ as it follows from Eq.\ (\ref{seq:refEquation12}), and in the entire range of $\Gamma$, $Z_i>0$. It is worth mentioning that at $\Gamma>300$, reflections of the cold ions from a spherical boundary are so rare that they can be neglected. Therefore, the difference between BOCP and Coulomb cluster diminishes, and for these systems, $Z_i$ must coincide. In \cite{Zhukhovitskii_2024}, a negative value $Z_i=-0.04\pm 0.06$ was obtained at $\Gamma=500$ as a result of the substantial simulation inaccuracy.

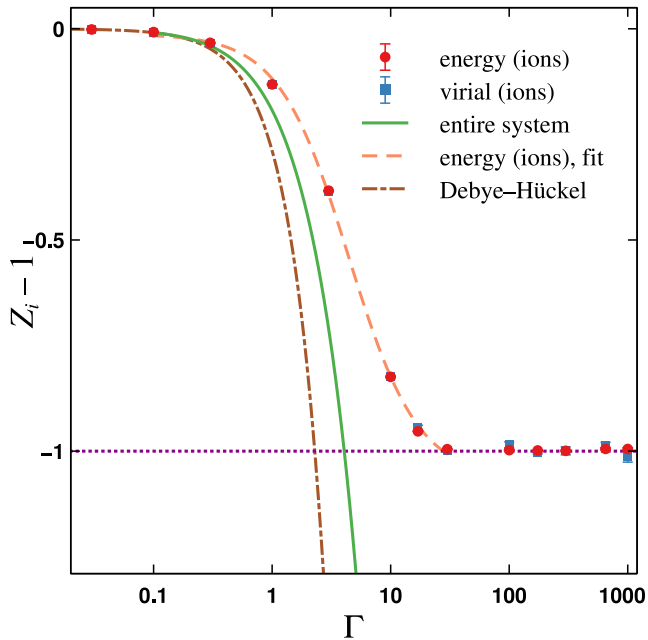
\begin{myfigure}{0.978}{9}{Ionic compressibility factor $Z_i$ determined from MD simulation using Eqs.\ (\ref{seq:refEquation15}) (circles) and (\ref{seq:refEquation23}) (squares). Dashed curve indicates the approximation (\ref{seq:refEquation42}) of $Z_i$ from the energies; solid line, the compressibility factor $Z$ for the entire system (\ref{seq:refEquation20}); and the dashed{}-dotted line, $Z$ in the Debye--Hückel approximation.}

\end{myfigure}

\section{Application to LAMMPS}
\subsection{Simulation procedure}
\label{bkm:RefNumPara55813304080641}A common way to calculate properties of an OCP system is to use the Ewald potential, which is typically split into two parts: $\text{erfc}(\alpha r)/r$, dominant at short and medium distances within the simulation cell, and $\text{erf}(\alpha r)/r$, a smooth, long-range function treated in reciprocal space. Here, $\alpha$ is the Ewald damping (splitting) parameter, chosen such that the real-space contribution at the cutoff length $r_c$ satisfies the condition $\text{erfc}(\alpha r_c)/r_c \lesssim \varepsilon_0$, where $\varepsilon_0$ is the target relative error for forces. Interactions within $r_c$ are calculated directly in real space, while interactions beyond $r_c$ are included via the reciprocal-space sum over periodic images. It is assumed that thermodynamic properties of the system are essentially independent of the precise cutoff, as long as $\alpha$ and $r_c$ are chosen to meet the desired accuracy $\varepsilon_0$.

To examine the effect of the cutoff length on derived thermodynamic quantities arising from specific computational algorithms, we performed a series of MD simulations using the stable release of LAMMPS (Large-scale Atomic/Molecular Massively Parallel Simulator, 29 August 2024) \cite{Thompson_2022} compiled with the ScaFaCoS library v1.0.4 \cite{scafacos_web}.

The total number of particles $N$ was fixed within each simulation, but varied between different simulations and the exact values will be specified later. Periodic boundary conditions were applied in all three directions of a cubic cell with the edge length $L=(4\pi N/3)^{1/3}$. The integration time step was chosen to be dependent on the Coulomb coupling parameter $\Gamma$ as $\tau = 10^{-3}(0.2\Gamma+2)$.

Ions of mass $m=1$ bearing the charge $ze=1$ were initialized randomly for fluid simulations and as a bcc lattice for solid-state simulations. Initial velocities were assigned the Maxwellian distribution corresponding to temperature $T = 1/\Gamma$. In both cases, the energies and forces were calculated by a direct summation in the real space inside a sphere of the radius $r_c$ around each ion and by summation in $k${}-space outside the sphere. Thus, long-range interactions were handled with the $k${}-space Ewald solver for simulations employing the Ewald method \cite{Karasawa_1989}, or with PPPM solver for simulations using the PPPM method \cite{Hockney_2021}; in both cases, with the target relative accuracy $\varepsilon_0 = 10^{-6}$. The Ewald method implies a direct calculation in the $k${}-space, while the PPPM method assigns charges to a finely-spaced mesh in the simulation box. Then the Poisson equation for the electrostatic potential is solved in the $k${}-space, and the force at each mesh point is calculated by numerically differentiating this potential. The force acting on an ion is evaluated as an interpolation of the mesh forces.

All simulations were carried out in the microcanonical ensemble. During the initial $5\times 10^6$ time steps, the Langevin thermostat with damping parameter 0.5 was applied. The thermostat was then disabled, and the system was allowed to evolve freely for at least $5\times 10^6$ additional time steps, while data were collected every 100 steps. The reported results were obtained by averaging over the final quarter of each trajectory.

\subsection{Correction to the calculation of ionic pressure}
\label{bkm:RefHeadingToc409121406508568}In order to demonstrate the dependence of the ionic compressibility factor of a system on the chosen cutoff length at fixed relative error, a series of MD simulations were performed under simulation conditions specified in Sec.~\ref{bkm:RefNumPara55813304080641}, while the total number of ions was chosen to be $N=10^3$ and the $k${}-space PPPM method was used.

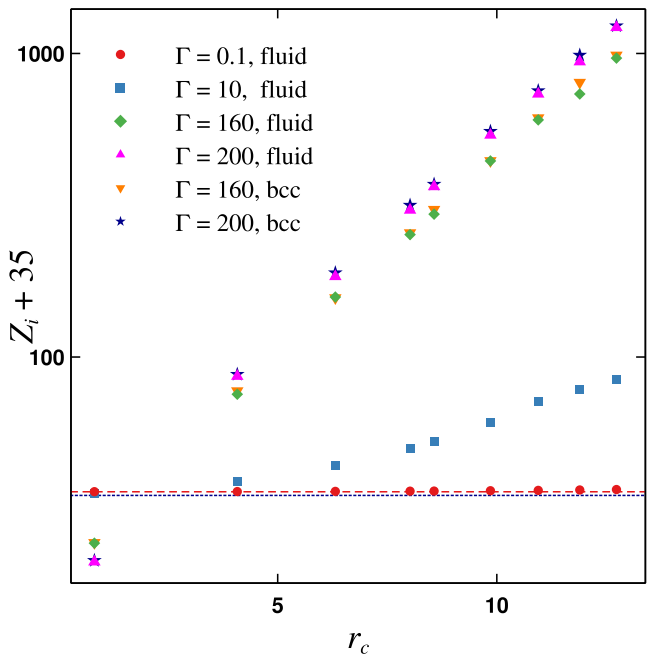
\begin{myfigure}{0.978}{10}{Ionic compressibility factor $Z_i$ for OCP as a function of the LAMMPS cutoff length $r_c$ for the cases of random initialization (fluid) at $\Gamma=0.1$ (circles), $\Gamma=10$ (squares), $\Gamma=160$ (diamonds), and $\Gamma=200$ (triangles up) and initialization as bcc (solid) at $\Gamma=160$ (triangles down) and $\Gamma=200$ (stars). Dashed and dotted lines show $Z_i$ obtained for BOCP at $\Gamma=0.1$ and $200$, respectively.}

\end{myfigure}

The ion compressibility factor $Z_i$ (\ref{seq:refEquation23}) shows a strong dependence on the cutoff length $r_c$, as illustrated in Fig.~\ref{f10} (a vertical shift of $Z_i$ by more than $14$ units is necessary to display it on a logarithmic scale), and this dependence becomes stronger with the increase in $\Gamma$ (cf.\ \cite{Onegin_2024}). In all cases, $Z_i$ grows as $r_c$ increases, indicating that the virial is also sensitive to $r_c$. For weak coupling ($\Gamma = 0.1$), $Z_i$ varies only slightly with $r_c$ and stays close to the BOCP reference value, demonstrating that the effect is present but relatively modest. At intermediate coupling ($\Gamma = 10$), the growth of $Z_i$ with $r_c$ becomes clearly visible though still moderate. In the strongly coupled regimes ($\Gamma = 160$ and $\Gamma = 200$), the dependence turns dramatic: $Z_i$ rises by orders of magnitude as $r_c$ increases. The choice of initial configuration (fluid vs.\ solid) appears to have no decisive influence on the resulting $Z_i$ values. A strong dependence of $Z_i$ on $r_c$ at high $\Gamma$ may be indicative of some artifact resulting, e.g., from a discontinuity of the effective interionic potential (and/or its spatial derivatives) at $r=r_c$ and the consequent singularity of its derivatives in coordinates at this point.

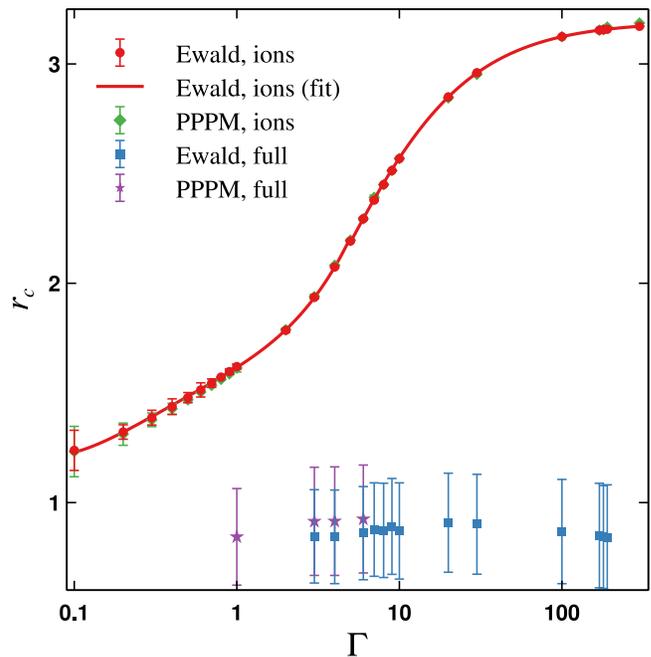
\begin{myfigure}{0.978}{11}{LAMMPS cutoff length $r_c$ ensuring equality of the compressibility factor calculated from the virial for OCP and the ionic compressibility factor of BOCP (\ref{seq:refEquation42}) for the Ewald and PPPM calculation methods (circles and diamonds, respectively), and equality of the compressibility factor calculated from the OCP virial and the OCP electrostatic energy (\ref{seq:refEquation20}) for the Ewald and PPPM methods (squares and stars, respectively). Solid line is the dependence $r_{c}(\Gamma)$ (\ref{seq:refEquation43}) and (\ref{seq:refEquation44}) fitting circles.}

\end{myfigure}

In this context, it becomes necessary to determine the proper cutoff length $r_c$ that yields the correct ionic pressure. This was done by numerically solving the equation $Z_{\mr{MD}}(r_c)=Z_i(\Gamma)$ for $r_c$ at the same $\Gamma$, i.e., by varying $r_c$ in the range from 0.1 to 16 and comparing the resulting $Z_{\mr{MD}}$ with the BOCP reference value $Z_i(\Gamma)$ (\ref{seq:refEquation42}). In all simulations, the total number of ions was fixed at $N=29^3$, and the ions were initialized randomly, except for the case $\Gamma=300$, where the bcc lattice initialization was used. We also attempted to determine the cutoff $r_c$ that ensures the correct compressibility factor of the entire system, by solving $Z_{\mr{MD}}(r_c)=Z(\Gamma)$ with $Z(\Gamma)$ defined by (\ref{seq:refEquation20}).

The resulting dependence of the optimal cutoff length $r_c$ on $\Gamma$ is shown in Fig.~\ref{f11}. For the ionic compressibility factor criterion, the cutoff obtained with the Ewald method exhibits a smooth monotonic increase with $\Gamma$, with the cutoff approaching $r_c \simeq 3$ at strong coupling. The PPPM method follows the same trend and shows no significant deviation from the Ewald results and also reduces the computational cost. The fitted curve is given by the expressions
\begin{equation}\label{seq:refEquation43}
r_{c}(\Gamma)=a_{0}+a_{1}\eta+a_{2}\eta^{2}+a_{3}\eta^{3}+a_{4}\eta^{4},
\end{equation}
where $\eta=\ln(\Gamma)$, $a_{0} = 1.61637$, $a_{1} = 0.21068$, $a_{2} = 0.035588$, $a_{3} = 0.024795$, and $a_{4} = 0.0078812$ at $0.1\le \Gamma \le 4$, and
\begin{equation}\label{seq:refEquation44}
r_{c}(\Gamma)={b_0}\tanh \left( {\frac{{\eta  + {b_1}}}{{{b_2}}}} \right) + {b_3},
\end{equation}
with $b_{0}=0.95012$, $b_{1}=-1.68645$, $b_{2}=1.7112$, and $b_{3}=2.23697$ at $4 < \Gamma \le 300$.

The compressibility factor of the entire system, which includes the neutralizing background, is lower than the ionic one, and therefore requires a significantly smaller cutoff length. However, for $r_c < 1$, the number of reciprocal lattice vectors required for the long-range part in reciprocal space, which is proportional to $\left[\ln(1/\varepsilon_0)/r_c\right]^3 N$, increases rapidly, leading to excessive memory usage. This memory limitation is the main source of large uncertainties obtained when attempting to determine the proper cutoff length for the full-system compressibility factor $Z$. The uncertainties shown in Fig.~\ref{f11} are calculated as the difference between the approximation of $r_c$ in the last successful run and the previous approximation. A failure in the determination of $r_c$ in the latter case confirms the main conclusion made in Sec.~\ref{bkm:RefNumPara97513304080641} that it is the ionic compressibility factor that is obtained using the virial from LAMMPS calculations rather than the compressibility factor for the entire system.

\section[Ambiguity of force{}-based quantities for the Coulomb OCP]{Ambiguity of force-based quantities for the Coulomb OCP}
\label{bkm:RefHeadingToc183112903921931}An explicit non-physical dependence of the ionic pressure on $r_c$ naturally leads to the hypothesis that existing convergence-acceleration schemes for the Coulomb sums reproduce the excess energy with high accuracy, but introduce non-negligible ambiguous distortions in the ionic forces. Once this force-level ambiguity is admitted, its consequences for kinetic and interfacial properties follow directly. For the fluid and solid OCP, the phonon spectra are determined by the dynamical matrix
\begin{equation}
D_{\alpha\beta}(i, j)=\frac{\partial^2 u}{\partial x_{i,\alpha}\partial x_{j,\beta}},
\end{equation}
where $x_{i,\alpha}$ is the $\alpha$ component of the coordinate of $i$th ion, i.e., by the second derivatives of potential, equivalently, by derivatives of the effective forces acting between the $i$th and $j$th ions. Even a small, systematic distortion of these forces may lead to a considerable error in the phonon frequencies, the group velocities, and, generally, in any quantity governed by the lattice dynamics. Similarly, the Green--Kubo expressions for transport coefficients involve microscopic fluxes that depend explicitly on forces. For example, the potential term of the virial stress tensor entering the Green--Kubo formulas for shear viscosity and thermal conductivity has the form
\begin{equation}
\sigma^{\mr{pot}}_{\alpha\beta}=\frac{1}{2V}\sum_{i\ne j}r_{ij,\alpha} f_{ij,\beta}.
\end{equation}
Even if the effective pair interionic potential reproduces energies to high accuracy, any systematic bias in forces leads to biased Green--Kubo integrals for the viscosity, electrical conductivity, and thermal conductivity.

The situation is even more delicate for interfacial properties, such as the surface tension $\gamma$, which is expressed in terms of the stress tensor $\sigma_{\alpha\beta}$ as
\begin{equation}\label{seq:refEquation47}
\gamma=\int_{-\infty}^{\infty}\left[\sigma_{nn}(z)-\sigma_{nn}^{\mr{bulk}}\right]dz,
\end{equation}
where $\sigma_{nn}^{\mr{bulk}}$ is the stress tensor component in the bulk phase and $z$ is the coordinate normal to the interface. In the classical OCP, for which no liquid--solid binodal exists, surface tension manifests itself only in metastable states involving crystallites in the fluid or fluid inclusions in the crystal. In such situations, $\gamma$ is determined entirely by subtle differences between the stress components in an inhomogeneous region, and it is therefore extremely sensitive to long-range force errors, even when the bulk energy is estimated almost perfectly. Since the Gibbs free energy barrier for the fluid--solid nucleation $\Delta G\sim\gamma$, the nucleation rate $J \sim \exp(-\Delta G/T)$ is very sensitive to $\gamma$. This, in turn, can result in a pronounced correlation between the width of a fluid--solid metastable region observed in MD simulations and $\gamma$, i.e., $r_c$.

In order to check this, the fluid--solid phase transition in the OCP was studied at $\Gamma=10\mbox{--}250$ with a step of 10. At each $\Gamma$, four simulations were performed: (i) $r_c=14$ with random (fluid) initialization, (ii) $r_c=14$ with the bcc lattice initialization, (iii) $r_c=r_c(\Gamma)$ with random initialization, where $r_c(\Gamma)$ is calculated using Eqs.\ (\ref{seq:refEquation43}) and (\ref{seq:refEquation44}), and (iv) $r_c=r_c(\Gamma)$ with the bcc lattice initialization. For simulations starting from the fluid phase, the number of ions was set to $29^3$, while for those initialized as a bcc lattice, $N=10^3$ was used.

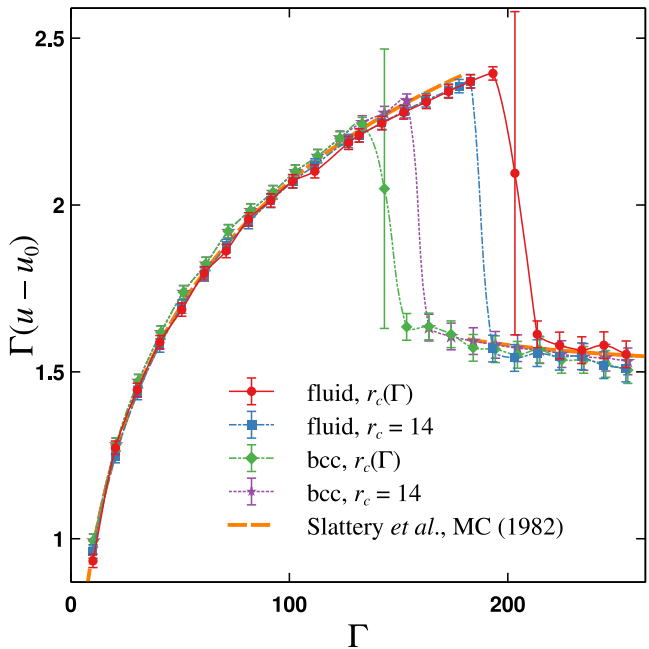
\begin{myfigure}{0.978}{12}{Thermal fraction of the potential energy as a function of the Coulomb coupling parameter for OCP. Circles and diamonds correspond to the simulation with $r_{c}(\Gamma)$ (\ref{seq:refEquation43}) and (\ref{seq:refEquation44}) and squares and stars, to $r_{c}=14$. The results from random (fluid, circles and squares) and bcc (solid, diamonds and stars) initialization are shown. Dashed line indicates curve fit of the results from study \cite{Slattery_1982}; lines connecting dots are guides for the eye.}

\end{myfigure}

The thermal fraction of the potential energy calculated from the system potential energy as $\Gamma(u-u_0)$ is shown in Fig.~\ref{f12}, where uncertainties are calculated as the difference between the averaged results of simulations performed under the same conditions but with different initial seeds of the random number generator. In a single-phase state, all simulation protocols reproduce the known dependence of $\Gamma(u-u_0)$ closely following the MC results by Slattery \textit{et al. }\cite{Slattery_1982}. The choice of different cutoff lengths ($r_c=14$ or $r_c(\Gamma)$) has little effect, and both the fluid and bcc initializations converge to almost identical values. This demonstrates that in the single-phase regime, the system equilibrates effectively, regardless of the initialization type and cutoff treatment. In contrast, differences become apparent near the fluid--solid transition. With the optimized cutoff $r_c(\Gamma)$, simulations initialized in the fluid state exhibit a sharp transition at $\Gamma=\mbox{180--200}$, while runs with fixed $r_c=14$ display the transition shifted to lower $\Gamma$, though following the same trend. For the bcc initialization, the system with $r_c=14$ undergoes the transition near $\Gamma \approx 150$, whereas those using the optimized $r_c(\Gamma)$ show the transition delayed to lower $\Gamma$. Overall, the metastable region of the OCP proves to be narrower when a fixed cutoff is used, while the optimized $r_c(\Gamma)$ yields a significantly broader metastability window. Such behavior can be rationalized in terms of the increasing dependence of the compressibility factor on the cutoff length in the vicinity of the optimized $r_c(\Gamma)$. Since $ \sum\nolimits_{i = 1}^N {\left\langle {{{\mathbf{r}}_i}{{\mathbf{f}}_i}} \right\rangle }<0$, it is seen from (\ref{seq:refEquation23}) and Fig.~\ref{f10} that a larger $r_c$ reduces the virial modulus thereby decreasing the surface tension $\gamma$ (\ref{seq:refEquation47}). A lower surface tension reduces the Gibbs free energy barrier for nucleation and enhances the nucleation rate thus accelerating the onset of phase transition and leading to the narrower metastable region at $r_c=14$, in accordance with our simulations.

Our findings indicate that standard Ewald-type treatments correspond to an OCP-like model, whose properties depend sensitively on how the calculation acceleration of the long-range Coulomb sums is controlled numerically, rather than the classical OCP. At the formal level, the exact Ewald summation corresponds to a unique OCP Hamiltonian and can, in principle, be evaluated to arbitrary precision. In actual MD simulations, however, one has to introduce finite real-space cutoffs, finite $k${}-space resolution, and target error tolerances, in addition to the unavoidable finite-size effects. A specific choice of $r_c$ may lead to sizeable, systematic distortions in forces and virial-based quantities such as the ionic compressibility factor. Hence, with respect to the kinetic and nonequilibrium properties, we are effectively testing an “Ewald-regularized OCP” with specific realization of the long-range summation algorithm. This motivates the development of methods that define the OCP EOS in terms of energy quantities only. The excess energy can be separated into the particle--particle and particle--background contributions like $u_{p\mathrm{ex}}$ and $u_{b\mathrm{ex}}$ and used to construct the corresponding ionic compressibility factor $Z_i^{(E)}(\Gamma)$ as well as related thermodynamic functions without any reference to the force virial. Apparently, this can be made by tuning the real-space cutoff $r_c$ or other realization-specific parameters. Then $Z_i^{(E)}$ could be used as a stringent consistency condition: only those implementations, for which the ionic compressibility factor obtained from forces matches $Z_i^{(E)}$ for the given $\Gamma$, should be trusted for the calculation of phonon spectra, transport coefficients, and interfacial properties.

\section{Conclusion}
In this study, we employed the model of BOCP to investigate the thermodynamic limit of a system of ions on the compensating background without introducing the PBC. This objective was reached due to the discovered universal form of the size dependence of system energy, which proved to be accurate enough to be extrapolated to an infinite number of the ions in a wide range of the Coulomb coupling parameter from 0.03 to 1000. This enabled us to avoid the Ewald summation procedure and estimate the characteristic energies of the system, which are free from additional strong interionic correlations imposed by the PBC. A very small width of the metastable region near the BOCP melting point makes it possible to evaluate its position, which is $\Gamma_{\mr m}=174\pm 2$. Also, we have introduced three converging quantities in addition to the system energy, namely, the excess interatomic electrostatic energy, the excess ion--background electrostatic energy, and the ionic pressure (compressibility factor), which we determined in the MD simulation. These quantities satisfy the limit conditions that follow from the theory. We show that neither the ionic pressure can be obtained from the energy of the entire system nor the excess pressure of the entire system can be estimated based on the virial determined from MD or MC simulations. We propose an accurate approximation of the ionic compressibility factor and compare it with that calculated for OCP using LAMMPS. We have established that at a certain value of the cutoff parameter $r_c$ introduced in LAMMPS, which is characteristic of the effective ionic interactions, the virial for OCP coincides with that for BOCP. We argue that this is a true virial ensured by a correct choice of $r_c$, which turns out to be a function of $\Gamma$. We present a wide-range approximation of this function for applications. Our data demonstrate that the system electrostatic energy is almost insensitive to $r_c$, while the ionic compressibility factor exhibits a strong dependence on this parameter. This indicates that convergence-acceleration schemes reproduce the energy landscape well but are much less accurate for the forces. Thus, the force-based quantities for the classical Coulomb OCP may become ambiguous.

Since $r_c$ defines the effective interionic forces, its correct choice is of crucial importance for the interfacial and dynamic properties of the Coulomb systems. As an example, we simulate the metastable region in the neighborhood of the system melting point with a true $r_c$ and the most commonly used. The results are indicative of the fact that for a true $r_c$, the width of the metastable region increases considerably, which points to the increase in the interfacial surface tension. We stress that in a correct classical OCP simulation, the obtained virial must be in compliance with the ionic compressibility factor listed in Table~\ref{t2} and shown in Fig.~\ref{f9}. Judging from Fig.~\ref{f10}, such compliance must be most important at high $\Gamma$.

Apart from a revision of our BOCP data for high $\Gamma$, an urgent problem to be addressed in the future is development of a calculation technique that would make it possible to determine the excess interatomic and ion--background energy directly in the simulation of OCP with PBC. Also, calculations of the ionic pressure tensor, phonon spectrum, and transport properties based on these energies are of undoubted interest.

\section*{Acknowledgments}
This work was supported by the Ministry of Science and Higher Education of the Russian Federation (State Assignment No. 075-00270-26-00).

\section*{AUTHOR DECLARATIONS}
\subsection*{Conflict of Interest}
The author has no conflicts to disclose.

\section*{DATA AVAILABILITY}
The data that support the findings of this study are available within the article.

\providecommand{\noopsort}[1]{} \providecommand{\singleletter}[1]{#1}%
\end{document}